\documentclass[prl,twocolumn,amsmath,amssymb,superscriptaddress]{revtex4-1}
\usepackage[utf8]{inputenc}
\usepackage[T1]{fontenc}
\usepackage{xcolor}
\usepackage{graphicx}
\usepackage{wasysym}
\usepackage{lmodern}
\usepackage{colonequals}
\usepackage{dcolumn}

\def\e{\mathrm{e}}
\def\I{\mathrm{i}}

\definecolor{darkblue}{rgb}{0,0,0.6}
\definecolor{darkred}{rgb}{0.6,0,0}
\usepackage[colorlinks=true,urlcolor=darkblue,citecolor=darkblue,linkcolor=darkred]{hyperref}

\usepackage{mathrsfs}

\usepackage{tikz}
\usepackage[kerning=true]{microtype}
\usetikzlibrary{patterns}

\newcommand*\circled[1]{\tikz[baseline=(char.base)]{
            \node[shape=circle,draw,inner sep=1.pt,minimum size=3ex] (char) {#1};}}

\newcommand{\dd}{\mathrm{d}}
\newcommand{\ed}{\mathrm{e}}

\DeclareMathOperator{\erfc}{erfc}

\def \equi#1{\mathrel{\mathop{\kern 0pt\sim}\limits_{#1}}}
\DeclareMathOperator{\erf}{erf}

\newcommand{\Df}{\mathcal{D}}

\def\O{\mathcal{O}}

\newcommand{\abs}[1]{\ensuremath{\left| #1 \right|}}
\newcommand{\moy}[1]{\ensuremath{\left\langle #1 \right\rangle}}

\newcommand{\ueta}{\underline{\eta}}

\begin{document}
\title{Exact closure and solution for spatial correlations in single-file diffusion}

\author{Aur\'elien Grabsch}
\affiliation{Sorbonne Universit\'e, CNRS, Laboratoire de Physique Th\'eorique de la Mati\`ere Condens\'ee (LPTMC), 4 Place Jussieu, 75005 Paris, France}

\author{Alexis Poncet}
\affiliation{Sorbonne Universit\'e, CNRS, Laboratoire de Physique Th\'eorique de la Mati\`ere Condens\'ee (LPTMC), 4 Place Jussieu, 75005 Paris, France}
\affiliation{Univ. Lyon, ENS de Lyon, Univ. Claude Bernard, CNRS, Laboratoire de Physique, F-69342, Lyon, France}

\author{Pierre Rizkallah}
\affiliation{Sorbonne Universit\'e, CNRS, Laboratoire de Physico-Chimie des \'Electrolytes et Nanosyst\`emes Interfaciaux (PHENIX), 4 Place Jussieu, 75005 Paris, France}

\author{Pierre Illien}
\affiliation{Sorbonne Universit\'e, CNRS, Laboratoire de Physico-Chimie des \'Electrolytes et Nanosyst\`emes Interfaciaux (PHENIX), 4 Place Jussieu, 75005 Paris, France}

\author{Olivier Bénichou}
\affiliation{Sorbonne Universit\'e, CNRS, Laboratoire de Physique Th\'eorique de la Mati\`ere Condens\'ee (LPTMC), 4 Place Jussieu, 75005 Paris, France}

\date{\today}

\begin{abstract}
Single-file transport, where particles diffuse in narrow channels while not overtaking each other, is a fundamental model for the tracer subdiffusion observed in confined systems, such as zeolites or carbon nanotubes. This anomalous behavior originates from strong bath-tracer correlations in 1D, which, despite extensive effort, have however remained elusive, because they involve an infinite hierarchy of equations. Here, for the Symmetric Exclusion Process, a paradigmatic model of single-file diffusion, we break the hierarchy and unveil a closed exact equation satisfied by these correlations, which we solve. Beyond quantifying the correlations, the central role of this key equation as a novel tool for interacting particle systems is further demonstrated by showing that it applies to out-of equilibrium situations, other observables and other representative single-file systems.
  \end{abstract}

\maketitle

\let\oldaddcontentsline\addcontentsline
\renewcommand{\addcontentsline}[3]{}

\section*{Introduction}

Single-file transport, where particles diffuse in narrow channels with the constraint that they cannot bypass each other, is a fundamental model~\cite{Levitt:1973,Arratia:1983} for tracer subdiffusion in confined systems. The very fact that the initial order is maintained at all times   leads to the subdiffusive behavior $\langle X_t^2\rangle \propto\sqrt{t}$ of the position of a tracer particle  (TP) \cite{Harris:1965}, in constrast with the regular  diffusion scaling $\langle X_t^2\rangle \propto{t}$. This theoretical prediction has been experimentally observed by microrheology in zeolites, transport of confined colloidal particles, or dipolar spheres in circular channels \cite{Hahn:1996,Wei:2000,Lin:2005}.

\begin{figure}
\centering
\includegraphics[width=0.49\textwidth]{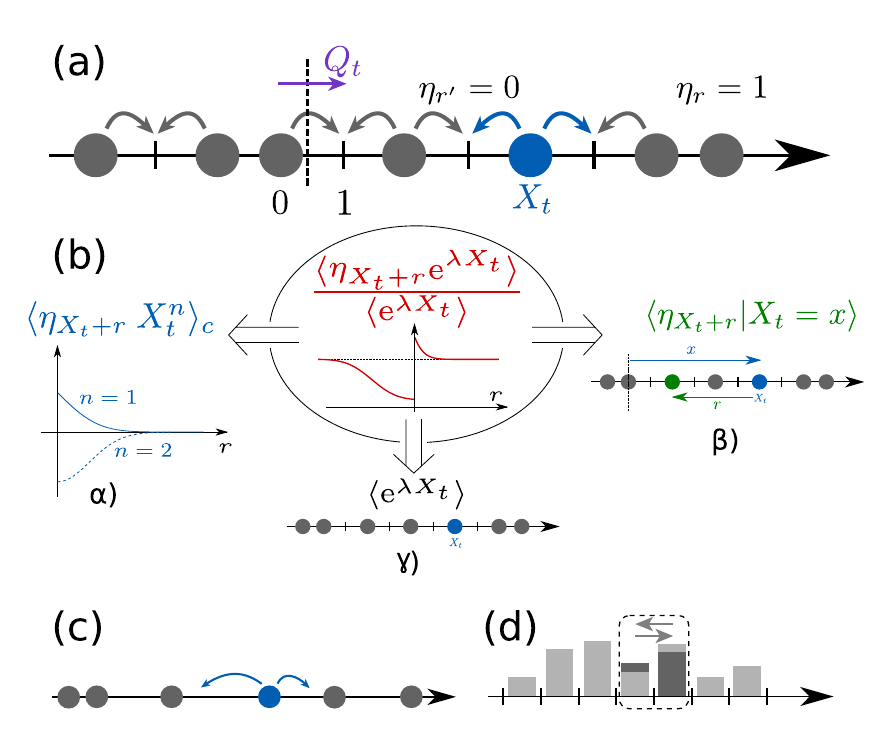}
\caption{ {\bf Models and bath-tracer correlations.} \textbf{(a)} The simple exclusion process (SEP). The tracer is located at position $X_t$, and $Q_t$ is the integrated current of particles through the origin up to time $t$ (i.e. the total flux of particles between sites $0$ and $1$). The occupation numbers of the sites are denoted $\eta_r$. \textbf{(b)} In this article, we put forward bath-tracer correlations $\langle \eta_{X_t+r} \mathrm{e}^{\lambda X_t} \rangle/ \langle \mathrm{e}^{\lambda X_t} \rangle$ (or bath-current correlations in the case of the integrated current $Q_t$) as fundamental quantities to analyze single-file diffusion, as they satisfy the simple closed integral equation~(\ref{eq:IntegEqOmP}). Besides being key technical tools, $\alpha$) they characterize the bath-tracer coupling (see Eq.~(\ref{scaling})), $\beta$) they quantify the response of the bath of particles to the perturbation induced by the tracer (see item (iv) after Eq.~(\ref{eq:psiPolyLogX})) and $\gamma$) in turn they control the large deviations of the subdiffusive motion of the tracer (see Eqs.~(\ref{eq:defPsi}), (\ref{eq:link_psi_w}) and~(\ref{eq:psiPolyLogX})).  Other representative models of single-file systems: \textbf{(c)} the Random Average Process: at exponential times, particles on a line can jump in either direction to a random fraction of the distance to the next particle; \textbf{(d)} the Kipnis-Marchioro-Presutti model: each site carries a continuous energy variable. At exponential times, the total energy of two neighbouring sites is randomly distributed between them.}
\label{fig:systems}
\end{figure}

The Symmetric Exclusion Process (SEP) is an essential model of  single-file diffusion. Particles, present at a density $\rho$, perform symmetric continuous-time random walks on a one-dimensional infinite lattice with unit jump rate, and with the hard-core constraint that there is at most one particle per site  (Fig.~1(a)). 
The SEP has become a paradigmatic model of statistical physics and it has generated a huge number of works in the mathematical and physical literature (see, e.g., Refs.~\cite{Spitzer:1970,Levitt:1973,Arratia:1983,Derrida:2009}). 
A major recent advance has been achieved with the calculation of the large deviation function of the position $X_t$ of a tracer in the long time limit \cite{Imamura:2017,Imamura:2021}. It gives access to all the long-time cumulants of $X_t$, which are in particular found to behave anomalously as $\sqrt{t}$ \cite{Krapivsky:2015a,Imamura:2017}.
 Similarly, the cumulants of the time integrated current through the origin $Q_t$ have been shown to also scale as  $\sqrt{t}$, and the large deviation function has been determined \cite{Derrida:2009}.

This collection of anomalous behaviors in the SEP originates from the  strong spatial correlations in the single-file geometry, which makes them determining quantities. Even if this has been recognized qualitatively for  long and that the case of dense and dilute limits have been recently studied \cite{poncet2021generalised}, up to now there is no  quantitative determination of the bath-tracer correlations at arbitrary density, despite extensive effort. Indeed, although the SEP has been studied for more than 40 years, analytic formulas for these functions are still missing. The calculation of these  correlations in the SEP actually constitutes an open many-body problem, which we solve here. More generally, we put forward  bath-tracer correlations as  fundamental quantities to analyze single-file diffusion, since we show that they satisfy a strikingly simple exact closed equation. The central role of this key equation as a novel tool for interacting particle systems is further demonstrated by showing that it applies to out-of equilibrium situations, other observables and other representative single-file systems.

We consider a SEP of average density $\rho$, with a tracer,  of position $X_t$  at time $t$, initially at the origin. The bath particles are described by the set of occupation numbers  $\eta_r(t)$ of each site $r\in\mathbb{Z}$ of the line at time $t$, with $\eta_r(t) = 1$ if the site is occupied and $\eta_r(t) = 0$ otherwise (see Fig.~1(a)). 
The statistics of the position of the tracer  is described by the cumulant-generating function, whose expansion defines the cumulants $\kappa_n$ of the position: 
\begin{equation}
\label{eq:defPsi}
\psi(\lambda, t) \equiv \ln \left\langle \mathrm{e}^{\lambda X_t} \right\rangle \equiv 
\sum_{n=1}^\infty \frac{\lambda^n}{n!} \kappa_n(t).
\end{equation}
Its evolution equation is given by (see Eq.~(S20) in Supplementary Information (SI)):
\begin{equation} \label{eq:link_psi_w}
\frac{\mathrm{d}\psi}{\mathrm{d} t} = \frac{1}{2} \left[ (\mathrm{e}^\lambda - 1)(1-w_1) + (\mathrm{e}^{-\lambda} - 1)(1-w_{-1}) \right],
\end{equation}
where the generalized density profiles (GDP) generating function  is defined by
\begin{equation}
\label{eq:def_gdp}
   w_r(\lambda, t) \equiv \langle \eta_{X_t+r} \mathrm{e}^{\lambda X_t}\rangle /\langle \mathrm{e}^{\lambda X_t}\rangle.
\end{equation}
Note that, besides controlling the time evolution of the cumulant-generating function,  $w_r$ (together with $\psi$) completely characterizes the joint cumulant-generating function of $(\eta_{X_t+r},\:X_t)$ and thus the bath-tracer correlations~\cite{poncet2021generalised}. The GDP-generating function is therefore a key quantity, and the next step consists in writing its evolution equation from the master equation describing the system. However, similarly to Eq.~(\ref{eq:link_psi_w}), it involves higher-order correlation functions. In fact, we are facing an infinite hierarchy of evolution equations, which is the rule for tracer diffusion (and for other observables such as the integrated current through the origin) in interacting particle systems \cite{Derrida_2007,Derrida:2004,P.L.-Krapivski:2009,Imamura:2017}, and whose closure has remained elusive up to now.  We provide below a closed equation which allows the determination of the GDP-generating function in the hydrodynamic limit (large time, large distance).

In this limit, the position of the tracer satisfies a large deviation principle~\cite{Sethuraman:2013,Imamura:2017,Imamura:2021}, which implies that the cumulant-generating function scales as $\psi\sim\hat{\psi}\sqrt{2t}$. In fact, this anomalous behavior originates from the more general scaling form
\begin{equation}
\label{scaling}
w_r(\lambda, t)  \mathop{\sim}\limits_{t \to \infty} \Phi\left(\lambda,v=\frac{r}{\sqrt{2t}}\right)\equiv\sum_{n=0}^\infty \frac{\lambda^n}{n!} \Phi_n(v)
\end{equation}
of the GDP-generating function, where the coefficient  $\Phi_n$ gives the large-scale limit of the  joint 
cumulant $\langle \eta_{X_t+r} X_t^n\rangle_c$ of the tracer's position $X_t$ and the occupation number $\eta_{X_t+r}$ measured in its frame of reference (Fig.~1(b)). In the following, we will drop the argument $\lambda$ of $\Phi$ for convenience.

\section*{Results}

We report here (see Materials and Methods and Section II.A of SI for details) that the two functions (rescaled derivatives of the profiles)
\begin{equation}
  \label{eq:defOmega}
  \Omega_\pm(v) \equiv \mp 2\hat{\psi} \frac{\Phi'(v)}{\Phi'(0^\pm)}
  \quad \mbox{defined for} \quad v \gtrless 0
\end{equation}
are entirely determined by the {\it closed} Wiener-Hopf integral equations~\cite{Polyanin:2008} with a Gaussian kernel
\begin{equation}
  \label{eq:IntegEqOmP}
  \Omega_\pm(v) = \mp \omega \: \mathrm{e}^{-(v+\xi)^2 + \xi^2}
  - \omega \int_{\mathbb{R}^\mp}  \Omega_\pm(z)  \mathrm{e}^{-(v-z+\xi)^2 + \xi^2}\mathrm{d} z,
\end{equation}
where $\xi \equiv \frac{\mathrm{d} \hat{\psi}}{\mathrm{d} \lambda}$ and we have analytically continued $\Omega_+$ to $v<0$ and
$\Omega_-$ to $v>0$.  The parameter $\omega$ is  determined by the boundary conditions (see Eq.~(\ref{eq:defOmega}))
\begin{equation}
  \label{eq:OmPMZeroPsi}
  \Omega_+(0) = -\Omega_-(0) = - 2 \hat{\psi},
\end{equation}
so that the functions $\Omega_\pm(v)$ are  parametrized by $\hat{\psi}$. 
At this stage, the expression of $\hat{\psi}(\lambda)$ has not been determined yet, but it can be obtained in the following way. First, $\Phi$ is deduced by integration of $\Omega_\pm$, with
\begin{equation}
  \label{eq:BoundInf}
  \Phi(\pm \infty) = \rho,
\end{equation}
by definition, and the boundary conditions 
\begin{equation}
  \label{eq:BoundDeriv}
  \Phi'(0^\pm) \pm 2 \frac{ \hat{\psi}}{\mathrm{e}^{\pm\lambda}-1} \Phi(0^\pm) = 0.
\end{equation}
The resulting  $\Phi$ are at this stage parametrized by $\hat{\psi}$ and $\lambda$. Then, by using the large time limit of Eq.~(\ref{eq:link_psi_w}),
\begin{equation}
\label{eq:Bound0}
  \frac{1-\Phi(0^-)}{1-\Phi(0^+)} = \mathrm{e}^{\lambda},
\end{equation}
 $\hat{\psi}$ can be written as a function of $\lambda$, and we finally  obtain the desired GDP-generating function $\Phi(\lambda,v)$.

\begin{figure*}
\centering
\includegraphics[width=\textwidth]{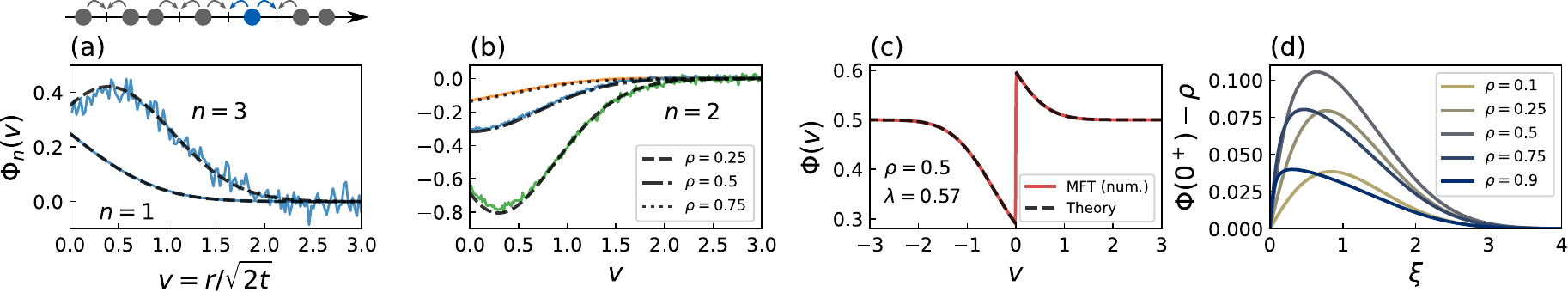}
\caption{\textbf{SEP}. Generalized density profiles (GDP) of order: \textbf{(a)} $n=1$ and $n=3$ at density $\rho=0.5$. Solid lines: numerical simulations at $t=3000$. Dashed lines: theoretical predictions (equations (S79,S80) in SI); \textbf{(b)} $n=2$ at densities $\rho=0.25$, $0.5$ and $0.75$. Solid lines: numerical simulations at $t=3000$. Dashed lines: theoretical predictions (equation (S79) in SI). Note that, interestingly $\Phi_2$ exhibits a minimum at a distance $v>0$ for $\rho<1/2$, which disappears for $\rho>1/2$. \textbf{(c)} GDP-generating function (or equivalently, conditional profiles) obtained from the theoretical prediction~\eqref{eq:IntegEqOmP} (dashed) compared to the numerical resolution of the MFT equations (solid line) described in Section IV.B in SI. \textbf{(d)} Non-monotony of the conditional profiles $\Phi(0^+)$ in front of the tracer, as a function of the rescaled tracer's position $\xi$, for different densities $\rho$.}
\label{fig:ProfSEP}
\end{figure*}

\section*{Discussion}
 
Several comments are in order. (i) We show in SI (Section I.E) that the boundary condition (\ref{eq:BoundDeriv}) is exact; furthermore,  we argue below that the bulk equation (\ref{eq:IntegEqOmP}), and thus the obtained GDP-generating function $\Phi$,  are also exact. (ii) Importantly, the Wiener-Hopf equations (\ref{eq:IntegEqOmP}) can be solved explicitly in terms of the one-sided Fourier transforms~\cite{Polyanin:2008}:
\begin{equation}
\label{eq:SolOmega}
    \int_{\mathbb{R}^\pm} \Omega_\pm(v) \mathrm{e}^{\mathrm{i} k v} \mathrm{d} v
    =
    \pm \left(
    1 - \exp[-Z_\pm]
    \right),
\end{equation}
where
\begin{equation}
\label{eq:SolOmegaZ}
    Z_\pm \equiv 
    \frac{1}{2} \sum_{n \geq 1}
      \frac{(-\omega \sqrt{\pi} \: \mathrm{e}^{-\frac{1}{4}(k+2\mathrm{i} \xi)^2})^n}{n}
      \: \mathrm{erfc} \left(\pm \sqrt{n} \left(\xi - \frac{\mathrm{i} k}{2} \right) \right).
\end{equation}
(iii) As a byproduct, our approach yields the cumulant generating function $\hat{\psi}$ (or equivalently the large deviation function of the tracer's position),
\begin{equation}
\label{eq:psiPolyLogX}
    \hat{\psi} =  -\frac{1}{2 \sqrt{\pi}}
  \mathrm{Li}_{\frac{3}{2}} (-\sqrt{\pi} \omega),
  \quad
  \mathrm{Li}_\nu(x) \equiv \sum_{n \geq 1} \frac{x^n}{n^\nu},
\end{equation}
which is shown in SI (Section II.E)  to be identical to the exact expression obtained using the arsenal of integrable probabilities in~\cite{Imamura:2017,Imamura:2021}.
(iv) Additionally, we obtain a full characterization of the spatial bath-tracer correlations and in particular analytical expressions of the $\Phi_n$ by using the procedure described above (see Section II.C of  SI for  explicit expressions,
which extend at arbitrary density the expressions given in~\cite{poncet2021generalised} in  the dilute ($\rho \to 0$) and dense ($\rho \to 1$) limits, and Fig.~2(a) and (b) for comparison with numerical simulations). 
(v) Our approach also provides the conditional profiles $\langle \eta_{X_t+r} | X_t=x \rangle$ defined as the average of the occupation of the site $X_t+r$ given that the tracer is at position $x$. Indeed, in the hydrodynamic limit, $\langle \eta_{X_t+r} | X_t=x \rangle \mathop{\sim}\limits_{t \to \infty} \Phi(\lambda^*,v)$ where $v = r/\sqrt{2t}$, $\lambda^*$ is defined by $\xi(\lambda^*) = \frac{\mathrm{d} \hat{\psi}}{\mathrm{d} \lambda}(\lambda^*) = x/\sqrt{2t}$ and $\Phi$ is the GDP-generating function determined above. While the (unconditional) profiles $\langle \eta_{X_t+r} \rangle$ are flat, the conditional profiles allow to probe the response of the bath of particles to the perturbation created by the displacement $\xi$ of the tracer: in particular, for $\xi >0$, it leads to an accumulation of bath particles in front of the tracer and a depletion behind (see Fig.~2(c)), quantified by the simple conservation relation
\begin{equation}
    \int_{0}^{+\infty} (\Phi(v)-\rho) \mathrm{d} v
    - \int_{-\infty}^0 (\Phi(v)-\rho) \mathrm{d} v
    = \rho \xi,
\end{equation}
which is a consequence of~(\ref{eq:SolOmega},\ref{eq:SolOmegaZ}). Another striking feature is the non-monotony of the conditional profile $\Phi(\lambda^*,0^+)$ in front of the tracer as a function of the rescaled position of the tracer $\xi(\lambda^*)$, see Fig.~2(d). Surprisingly, $\Phi(\lambda^*,0^+)$ does not saturate to $1$ as $\xi \to +\infty$, but instead returns to its unperturbed value $\rho$, which results from a global displacement of bath particles induced by the tracer.

\begin{figure*}
\centering
\includegraphics[width=\textwidth]{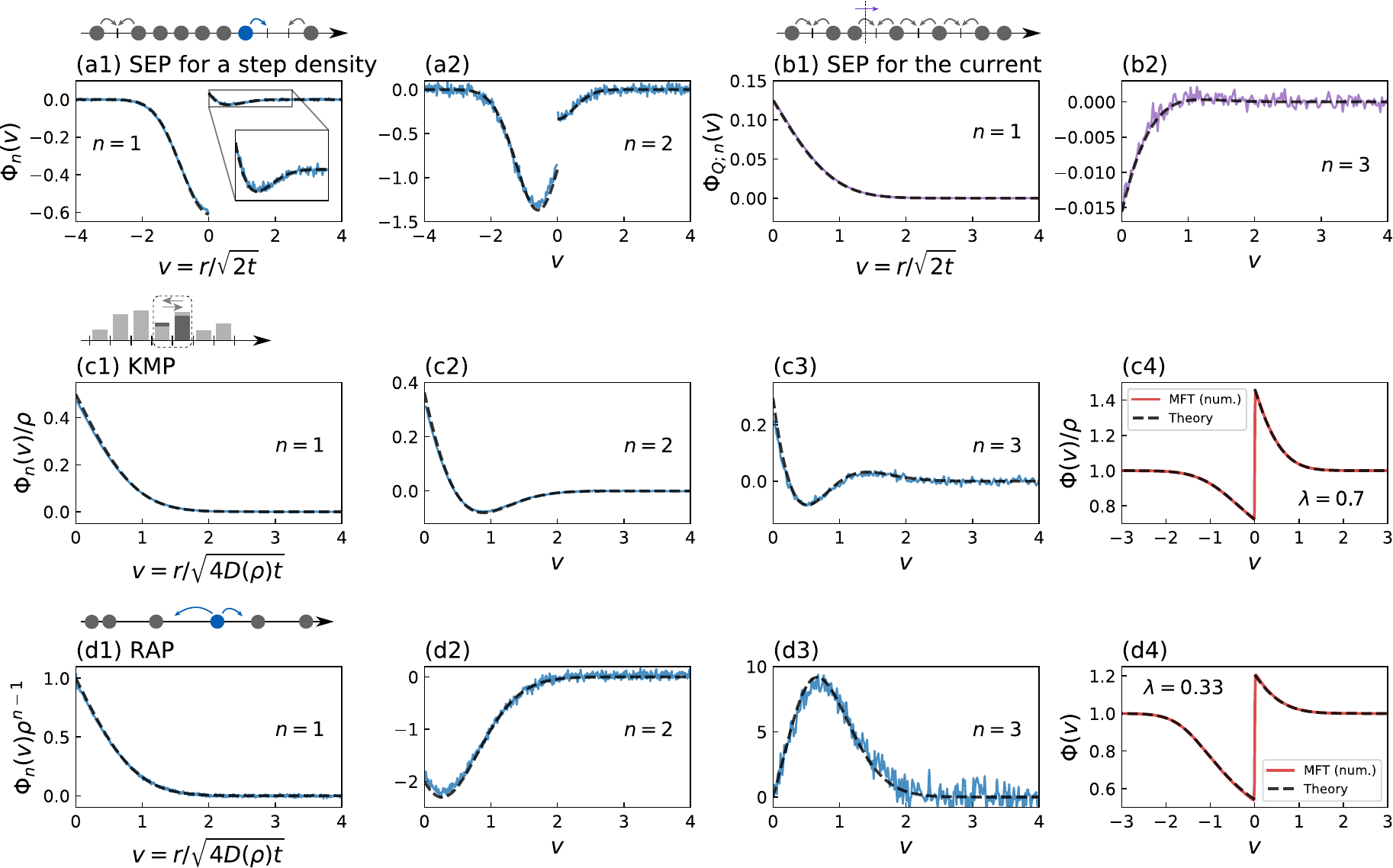}
\caption{\textbf{Extensions:} \textbf{(a)} The out of equilibrium situation of an initial step density. \textbf{(b)} Another observable, the current through the origin. \textbf{(c),(d)} Other representative single-file systems: the KMP model (c) and the RAP (d).
The dashed lines correspond to the predictions obtained from the central equation~\eqref{eq:IntegEqOmP}. (a) GDPs at order $n=1$ and $n=2$ for the SEP with a step initial density $\rho_-=0.7$ and $\rho_+=0.2$ at $t=1500$ with $2000$ sites. (b) GDPs for the current $\langle \eta_r Q_t^n \rangle_c$ in the SEP for a density $\rho=0.5$ at $t=900$, for orders $n=1$ and $n=3$ (the profile for $n=2$ is zero).
(c) KMP model ($D(\rho)=1/2$ and $\sigma(\rho) = \rho^2$) for $\rho=1$. (c1)-(c3) First three orders of the GDPs at $t=900$ and $500$ sites. (c4) GDP-generating function  (d) RAP for $\rho=1$ (for a uniform jump distribution, $D(\rho)=1/(4\rho)$ and $\sigma(\rho) = 1/\rho$). (d1)-(d3) First three orders of the GDPs at $t=4000$ with $5000$ particles (solid lines). (d4) GDP-generating function compared to the numerical resolution of the MFT equations (solid line).}
\label{fig:ExtenProf}
\end{figure*}

Importantly, Equation~(\ref{eq:IntegEqOmP}) describes several other situations of physical relevance. 
(i) First, it applies to the out-of-equilibrium situation of an initial step of density $\rho_+$ for $x>0$ and $\rho_-$ for $x<0$, with the tracer initially at the origin. This paradigmatic setup has attracted a lot of attention \cite{Derrida:2009,Derrida:2009a,Krapivsky:2012,Imamura:2017,Imamura:2021} since it remains transient at all times and never reaches a stationary state. The GDP-generating function $\Phi$ is obtained from the solution~(\ref{eq:SolOmega}) by following the procedure described above, upon only changing the boundary condition~(\ref{eq:BoundInf}) into $\Phi(\pm \infty) = \rho_\pm$. Again, we recover the results of~\cite{Imamura:2017,Imamura:2021} on the cumulant generating function $\hat{\psi}$. Additionally, we obtain the complete spatial structure of the bath-tracer correlations (see Fig.~3(a) and Section II.D.1 of SI for explicit expressions).
(ii) Second, and strikingly, it also gives access to the statistics of other observables, as exemplified by the case of the integrated current through the origin $Q_t$ (see Section II.D.3 of  SI for the application to the generalized current, which is an extra observable), defined as the total flux of particles between sites $0$ and $1$ during a time $t$. This quantity has been the focus of many studies, both in the context of statistical physics~\cite{Spohn:1989,Derrida_2007,Derrida:2009,Derrida:2009a,PhysRevE.101.052101} and mesoscopic transport~\cite{Beenakker:1992,BLANTER:2000,Lee:1995}, in particular in the nonequilibrium situation $\rho_- \neq \rho_+$ \cite{Derrida:2009,Derrida:2009a}.
Note that while the statistics of tracer diffusion and integrated current are easily related in the case of quenched initial conditions~\cite{Sadhu:2015}, the relation is more entangled for the annealed case considered here due to the fluctuations of the initial condition. 
The quantities introduced previously~(\ref{eq:defPsi}-\ref{eq:def_gdp}) on the example of tracer diffusion are naturally adapted by substituting $Q_t$ for $X_t$. The corresponding profiles $\Phi_Q$ are then obtained as a particular case of Equation~(\ref{eq:IntegEqOmP}) by setting $\xi=0$, completed by modified boundary conditions~(\ref{eq:BoundDeriv},\ref{eq:Bound0}) derived from the microscopic model (see Section II.D.2 of SI). In particular, the resulting Eq.~(\ref{eq:psiPolyLogX}) gives back the exact cumulant generating function of $Q_t$ obtained in \cite{Derrida:2009} by Bethe ansatz, since in this case we find that
\begin{equation}
\omega_Q \sqrt{\pi} = \rho_- (1-\rho_+) (\mathrm{e}^\lambda - 1) + \rho_+ (1-\rho_-) (\mathrm{e}^{-\lambda} - 1),
\end{equation}
which coincides with the single parameter involved in~\cite{Derrida:2009,Derrida:2009a}. Additionally, the $\Phi_Q$ determined here provides the associated spatial structure (see Fig.~3(b) and Section II.D.2 of SI for explicit expressions). These profiles have been introduced and studied numerically in \cite{Gerschenfeld:2012} for an infinite system (see also \cite{Derrida:2004} for a finite system between two reservoirs), but no analytical expressions were available until now.
(iii) Finally, beyond the SEP, it applies to other representative single-file systems of interacting particles with average density $\rho$ (see also~\cite{poncet2021generalised}, which is however limited to the calculation of the first order $\Phi_1$). Such systems can be described at large scale by two quantities: the diffusivity $D(\rho)$ and the mobility $\sigma(\rho)$~\cite{Spohn:1991}. The case of the SEP considered above corresponds to $D(\rho) = 1/2$ and $\sigma(\rho) = \rho(1-\rho)$. Equation~(\ref{eq:IntegEqOmP}) (with adaptations of equations~(\ref{eq:BoundDeriv},\ref{eq:Bound0}) given in Section III.C of SI) more generally applies to single-file systems with $D(\rho) = 1/2$, $\sigma''(\rho)$ constant and $\sigma(0)=0$, by replacing the $-\omega$ that multiplies the integral in~(\ref{eq:IntegEqOmP}) by $\omega \sigma''(\rho)/2$.
Important cases covered by our approach include (see Fig.~1 for definitions and Sections III.D and III.E of SI for explicit expressions): (a) the model of hard Brownian particles ($\sigma(\rho) = \rho$) for which the GDP-generating function of~\cite{poncet2021generalised} is recovered; (b) the Kipnis-Marchioro-Presutti (KMP) model~\cite{Kipnis:1982} ($\sigma(\rho) =\rho^2$, see Fig.~3(c)), which describes situations as varied as force fluctuations in packs of granular beads~\cite{Liu:1995}, the formation of clouds and gels, self-assembly of molecules in organic and inorganic materials and distribution of wealth in a society (see~\cite{Das:2017} and references therein). (c) the Random Average Process (RAP), which appears in a variety of problems such as force propagation in granular media, models of mass transport or models of voting systems~\cite{Liu:1995,10.1214/EJP.v3-28,Krug:2000,Rajesh:2000}. Although $\sigma''(\rho)$ is not constant in this case, the GDP-generating function can be deduced from our results thanks to a mapping between tracer diffusion in the RAP and the integrated current in the KMP model~\cite{Kundu:2016} (see Fig.~3(d)).

We finally argue that the central equation~(\ref{eq:IntegEqOmP}) is exact for the following reasons. (i) We show in SI (Section IV) that the Macroscopic Fluctuations Theory (MFT)~\cite{Bertini:2015} can be used to determine perturbatively the first coefficients $\Phi_n$ analytically. These coefficients computed up to order $n=5$ (which is the highest order for which we managed to determine the integrals involved) coincide with those obtained by our approach. Furthermore, the agreement holds also non-perturbatively in $\lambda$, as displayed in Fig.~2(c) and Fig.~3(c4),(d4) where the numerical solution of the MFT equations is compared to the analytical solution~(\ref{eq:SolOmega},\ref{eq:SolOmegaZ}).
Moreover, and as mentioned above, the exact expression of the cumulant generating functions of (ii) the tracer position of~\cite{Imamura:2017} and (iii) the integrated current of~\cite{Derrida:2009} are contained in our approach, including the case of an initial step of density.

All together, we have  determined analytically the spatial correlations in the SEP, which  allowed us to fully quantify the response of the bath to the perturbation induced by a tracer. Besides being paramount physical observables,  these  correlations have been shown to be  fundamental technical quantities, since they satisfy a strikingly simple closed equation and control large deviations in single-file diffusion. This very same equation  applies to a variety of situations involving single-file transport, which makes it a novel and promising tool to tackle interacting particle systems.

\section*{Materials and Methods}

\subsection*{Analytical calculations for the SEP}

Details on analytical calculations are provided in SI. We sketch here the main steps that led to the closed equation~(\ref{eq:IntegEqOmP}) for the SEP. The starting point is a master equation describing the time evolution of the complete system (bath and tracer in the SEP), from which we obtain the time evolution of the cumulant generating function $\psi$ and the GDP-generating function~(\ref{eq:def_gdp}). The main difficulty is that the latter involves higher-order correlation functions.

The next step consists in using the scaling~(\ref{scaling}) of the GDP-generating function and $\psi(\lambda,t) \sim \hat\psi(\lambda) \sqrt{2t}$ to derive the hydrodynamic limit of the problem (details given in Section I.E of SI). The obtained bulk equation, valid at arbitrary density, is still not closed. We explain in SI that a closed equation obeyed by $\Phi$ has to satisfy the following constraints: (i) it must reduce to the known equations obtained in the limits of high and low density in~\cite{poncet2021generalised}; (ii) it should also reproduce, as a byproduct, the cumulants of the tracer's position derived recently in~\cite{Imamura:2017,Imamura:2021}; (iii) additional constraints concern the way the different parameters appear in the equation (see Section II.A of SI for details); (iv) finally, the equation we write should have a "proper scaling" with time determined in Section I.E of SI.

Following these ideas and constraints, we obtain a first closed equation which holds at lowest orders in $\lambda$, see (S40) in SI, which properly reproduces the known cumulant $\kappa_n$ for $n \leq 6$. This equation is conveniently rewritten by introducing the new functions $\Omega_\pm(v)$ defined by Eq.~(\ref{eq:defOmega}). Extension of this equation to arbitrary order in $\lambda$ and then its resummation yields the closed equations~(S48,S49) of SI. Several technical steps detailed in Section II.A of SI allow us to finally transform them into the closed Wiener-Hopf integral equation~(\ref{eq:IntegEqOmP}), which is our central result.

\subsection*{Numerical simulations for the SEP}

The numerical simulations of the SEP are performed on a periodic ring of size $N$, with $M=\rho N$ particles at average density $\rho$. The particles are initially placed uniformly at random. The jumps of the particles are implemented as follow: one picks a particle uniformly at random, along with one direction (left and right with equal probabilities). If the chosen particle has no neighbor in that direction, the jump is performed, otherwise it is rejected. In both cases, the time of the simulation is incremented by a random number picked from an exponential distribution of rate $N$. We keep track of one particle (the tracer) and compute the moments of its displacement and the generalized density profiles. The averaging is performed over $10^8$ simulation.

\subsection*{Extensions}

Analytical (Section III of SI) and numerical (Section V of SI) extensions (other systems than the SEP, other observables, nonequilibrium situations), following these lines, are described in SI.

\bibliographystyle{apsrev4-1}

\let\addcontentsline\oldaddcontentsline

\clearpage
\widetext

\begin{large}
  \begin{center}

    \textbf{
      Exact closure and solution for spatial correlations in single-file diffusion
    }

    \bigskip

    \textbf{Supplementary Information}

  \end{center}
\end{large}

\setcounter{secnumdepth}{3}

\renewcommand{\theequation}{S\arabic{equation}}
\renewcommand{\thefigure}{S\arabic{figure}}

\setcounter{equation}{0}

\tableofcontents

\section{General equations and hydrodynamic limit}

In this Section, we mostly recall the equations and results obtained in~\cite{poncet2021generalisedSI} in order to provide a self-contained document. We additionally introduce in Section~\ref{sec:CondProf} the conditional profiles discussed in the main text after Eq.~(13). 

\subsection{Master equation of the SEP}

We consider the symmetric exclusion process (SEP) with a tracer at position $X$. We denote the configuration of the system by $\ueta = \{\eta_r\}_{r\in\mathbb{Z}}$ with $\eta_r \in {\{0, 1\}}$ the occupation number of site $r$ by the bath particles ($\eta_r = 1$ if site $r$ is occupied, $0$ otherwise). At time $t$, the system is characterized by a probability law $P(X, \ueta, t)$.

We initially start from the equilibrium distribution of the occupations, and the tracer at the origin (with the convention that the site occupied by the tracer is empty of bath particles):
\begin{equation}
    P(X, \ueta, 0) =
    \delta_{X, 0} \delta_{\eta_0, 0} \prod_{r\in\mathbb{Z}^\ast} \delta_{\eta_r, \gamma_r}
    \:,
\end{equation} 
where $\gamma_r$ are independent Bernouilli variables with parameter $\rho$ (density of the system) { and the site $0$ is treated independently because it is occupied by the tracer, and not a bath particle}.

The probability $P(X, \ueta, t)$ obeys the following master equation,
\begin{align} \label{eq:sm_master}
    \partial_t P(X, \ueta, t) 
    =&~ \frac{1}{2}\sum_{r\neq X, X-1} \left[P(X, \ueta^{r,+}, t)-P(X, \ueta, t)\right] \nonumber 
    \\
    &+ \frac{1}{2} \sum_{\mu=\pm 1} \left\{ (1 - \eta_X) P(X-\mu, \ueta, t) 
    - (1-\eta_{X+\mu}) P(X, \ueta, t) \right\} 
    \:,
\end{align}
where $\ueta^{r,+}$ is the configuration $\ueta$ in which the occupations of sites $r$ and $r+1$ are exchanged. The first term corresponds to the jumps of the bath particles while the second one takes into account the displacement of the tracer. 

\subsection{Observables and large-times scalings}
\label{ss:sm_obs}

We consider the cumulant-generating function of the displacement of the tracer,
\begin{equation}
    \label{eq:sm_def_psi}
    \psi(\lambda, t) \equiv \ln \left\langle \e^{\lambda X_t} \right\rangle.
\end{equation}
At large time $t$, it scales as $\sqrt{t}$~\cite{Imamura:2017SI,Imamura:2021SI},
\begin{equation}
    \label{eq:DefPsiAndExpLamb}
    \psi(\lambda, t) \equi{t\to\infty} \hat{\psi}(\lambda)\sqrt{2t}
    \:,
    \quad
    \hat{\psi}(\lambda) = \sum_{n \geq 1} \kappa_n \frac{\lambda^n}{n!}
    \:,
\end{equation}
where $\kappa_n$ is the $n^{\mathrm{th}}$ cumulant of the tracer's position (rescaled by $\sqrt{2t}$).
We also consider the generalized profiles,
\begin{equation}
    \label{eq:sm_def_wr}
    w_r(\lambda, t) \equiv 
    \frac{\langle \eta_{X_t+r} \e^{\lambda X_t}\rangle}{\langle \e^{\lambda X_t}\rangle}
    \:.
\end{equation}
Their expansion in powers of $\lambda$ gives the cross-cumulants $\moy{\eta_{X_t+r}\: X_t^n}_c$ between the occupations and the position of the tracer. At large time, they satisfy a diffusive scaling,
\begin{equation} 
    \label{eq:sm_scale_w}
    w_r(\lambda, t) \equi{t\to\infty}  \Phi\left(v = \frac{r}{\sqrt{2t}}, \lambda\right)
    \:,
    \quad
    \Phi(v, \lambda) = \sum_{n \geq 1} \Phi_n(v) \frac{\lambda^n}{n!}
    \:.
\end{equation}
{ This scaling is based on observations originating from numerical simulations. In addition it is compatible with the known scaling of the cumulant generating function~\eqref{eq:DefPsiAndExpLamb}. }

Finally, we consider the ``modified centered correlations'',
\begin{equation} \label{eq:sm_def_f}
    f_{\mu, r}(\lambda, t) \equiv \displaystyle
    \frac{\left\langle(1-\eta_{X_t+\mu})\eta_{X_t+r}\e^{\lambda X_t}\right\rangle}{\langle \e^{\lambda X_t}\rangle} - 
    \begin{cases}
        (1-w_\mu) w_{r-\mu} & \text{ if } \mu r > 0 \:, \\
        (1-w_\mu) w_r  & \text{ if } \mu r < 0 \:,
    \end{cases}
\end{equation}
At large time, the leading term is in $t^{-1/2}$ and the sub-leading term in $t^{-1}$ with the same diffusive scaling as for the profiles,
\begin{equation} \label{eq:sm_scale_f}
    f_{\mu, r}(\lambda, t) 
    = \frac{1}{\sqrt{2t}} F_\mu\left(v=\frac{r}{\sqrt{2t}}, \lambda\right)
    + \frac{1}{2t} G_\mu\left(v=\frac{r}{\sqrt{2t}}, \lambda\right) + \mathcal{O}(t^{-3/2})
    \:.
\end{equation}
{ This scaling ansatz relies again on numerical observations.}

\subsection{Equivalent description: large deviations and conditional profiles}
\label{sec:CondProf}

Alternatively, we can also consider the distribution of the position $X_t$ of the tracer at time $t$, which satisfies a large deviation principle~\cite{Sethuraman:2013SI,Imamura:2017SI,Imamura:2021SI},
\begin{equation}
  \label{eq:largeDevProb}
  \mathbb{P}(X_t = x)
  \underset{t \to \infty}{\simeq}
  \e^{-\sqrt{2t} \phi(\xi)}
  \:,
  \quad
  \xi = \frac{x}{\sqrt{2t}}
  \:,
\end{equation}
where $\psi$ is the large deviation function. The moment generating function of the position of the tracer is the Laplace transform of the distribution. We can show that these functions are simply related by writing
\begin{equation}
  \moy{\e^{\lambda X_t}} = \sum_x \e^{\lambda x} \mathbb{P}(X_t = x)
  \:.
\end{equation}
Using the scaling form~(\ref{eq:largeDevProb}), and taking the continuous limit, we have the following integral representation for large $t$:
\begin{equation}
  \moy{\e^{\lambda X_t}} \simeq
  \int \dd \xi \: \e^{\sqrt{2t} (\lambda \xi - \phi(\xi))}
  \:.
\end{equation}
The integral can be estimated with a saddle point approximation, which gives
\begin{equation}
  \moy{\e^{\lambda X_t}} \simeq
  \e^{\sqrt{2t} (\lambda \xi_* - \phi(\xi_*))}
  \:,
  \quad
  \phi'(\xi_*) = \lambda
  \:.
\end{equation}
The two functions $\phi$ and $\hat{\psi}$ are thus related by a Legendre transform
\begin{equation}
  \label{eq:RelPsiPhi}
  \hat{\psi}(\lambda) = \lambda \xi_*(\lambda) - \phi(\xi_*(\lambda))
  \:,
  \quad
  \phi'(\xi_*(\lambda)) = \lambda
  \:,
\end{equation}
or equivalently,
\begin{equation}
  \label{eq:RelPhiPsi}
  \phi(\xi) = \lambda_*(\xi) \xi - \hat{\psi}(\lambda_*(\xi))
  \:,
  \quad
  \hat{\psi}'(\lambda_*(\xi)) = \xi
  \:.
\end{equation}

In this language, it is natural to introduce the conditional profiles
\begin{equation}
  \label{eq:DefCondProf}
    \moy{\eta_{X_t+r} | X_t = x } = \mathbb{P}(\eta_{X_t+r}=1 | X_t = x)
\end{equation}
which give the probability that a site located at a distance $r$ from the tracer is occupied, given that the tracer is located at $x$. We can show the equivalence between the generalised profiles~(\ref{eq:sm_def_wr}) and the conditional profiles~(\ref{eq:DefCondProf}) by writing
\begin{equation}
  w_r(\lambda,t)
  = \frac{\moy{\eta_{X_t+r} \e^{\lambda X_t}}}{\moy{ \e^{\lambda X_t} }}
  = \frac{
    \displaystyle
    \sum_x \e^{\lambda x} \mathbb{P}(\eta_{X_t+r}=1| X_t = x) \mathbb{P}(X_t = x)
  }
  {\displaystyle \sum_x \e^{\lambda x} \mathbb{P}(X_t = x)}
  \:.
\end{equation}
Defining the large $t$ asymptotic form of the conditional profiles as
\begin{equation}
  \mathbb{P}(\eta_{X_t+r}=1 | X_t = x)
  \underset{t \to \infty}{\simeq}
  \tilde{\Phi} \left(\xi = \frac{x}{\sqrt{2t}}, v = \frac{r}{\sqrt{2t}} \right)
  \:,
\end{equation}
we obtain
\begin{equation}
  w_r(\lambda,t) \simeq \Phi(\lambda,v)
  = \frac{
    \displaystyle \int \dd \xi \:
    \tilde{\Phi}(\xi,v)
    \: \e^{\sqrt{2t} (\lambda \xi - \phi(\xi))}
  }
  {
    \displaystyle \int \dd \xi \: \e^{\sqrt{2t} (\lambda \xi - \phi(\xi))}
  }
  \simeq \tilde{\Phi}(\xi_*(\lambda),v)
  \:,
\end{equation}
or equivalently,
\begin{equation}
  \label{eq:RelPhiTPhi}
  \tilde{\Phi}(\xi, v) = \Phi(\lambda_*(\xi),v)
  \:.
\end{equation}
These two functions being equivalent, we will drop the variables $\lambda$ and $\xi$ and simply denote them both by $\Phi(v)$.

\subsection{Equations at arbitrary time}
\label{ss:sm_eqs}

Using Eqs.~\eqref{eq:sm_master}, one obtains the following equations for the time-evolution of the cumulant-generating function and of the generalized profiles.
\begin{align}
    \label{eq:sm_evolPsi}
    \partial_t \psi 
    &= \frac{1}{2}\left\{ (\e^\lambda-1)(1-w_1) + (\e^{-\lambda}-1)(1-w_{-1}) \right\}
    \:, \\
    \label{eq:sm_decoup_1}
    \partial_t w_r 
    &= \frac{1}{2} \Delta w_r - B_\nu \nabla_{-\nu}w_r 
    + \frac{1}{2} \sum_{\mu=\pm 1} 
    \left( \e^{\mu\lambda} f_{\mu, r+\mu} - f_{\mu, r} \right) \qquad (r\neq \pm 1) 
    \\
    \label{eq:decoup_2}
    \partial_t w_{\pm 1} 
    &= \frac{1}{2} \nabla_\pm w_{\pm 1} + B_{\pm} w_{\pm 1} 
    + \frac{1}{2} \left( \e^{\pm\lambda} f_{\pm 1, \pm 2} - f_{\mp 1, \pm 1} \right)
\end{align}
where $\nu$ is the sign of $r$, the gradients are $\nabla_\mu u_r = u_{r+\mu} - u_r$, {$\Delta u_r = u_{r+1} -2 u_r + u_{r-1}$} and
\begin{equation}
    B_{\pm} = \frac{\partial_t \psi}{\e^{\pm\lambda} - 1}
    \:.
\end{equation}
In addition, the generalized profiles at large distance are equal to the density: $\lim_{r\to\pm\infty}w_r = \rho$.

\subsection{Hydrodynamic equations at large time} \label{ss:sm_eqs2}

Using the scalings of Section~\ref{ss:sm_obs} into the {equation~(\ref{eq:sm_evolPsi}), we first obtain at order $0$ in $t$,
\begin{equation}
    (\e^\lambda -1) (1-\Phi(0^+)) + 
    (\e^{-\lambda} -1) (1-\Phi(0^-))
    = 0 \:,
\end{equation}
which we can rewrite as
\begin{equation}
\label{eq:sm_decoup_3b}
    \frac{1-\Phi(0^-)}{1-\Phi(0^+)} = 
    -\frac{\e^\lambda - 1}{\e^{-\lambda}-1}
    = \e^{\lambda}
    \:.
\end{equation}
Similarly, using these same scalings into~\eqref{eq:sm_decoup_1}, we first obtain at order $1/\sqrt{t}$,
\begin{equation}
    F_{-1}(v) = \e^\lambda F_1(v)
    \:,
\end{equation}
as well as the following hydrodynamic equations for the generalized profiles at order $1/t$,}
\begin{align}
    \label{eq:sm_decoup_1b}
    &\Phi''(v) + 2(v+ b_\nu) \Phi'(v) + C(v) = 0 
    \:, \qquad
    b_\nu = \nu \frac{\hat{\psi}}{\e^{\nu \lambda} - 1}
    \\
    \label{eq:exprCv}
    &C(v) = (\e^\lambda - 1)F_1'(v) + \sum_{\mu=\pm 1} (\e^{\mu\lambda}-1) G_\mu(v)
    \:,
\end{align}
with $\nu$ the sign of $v$. { Finally, using the scalings of Section~\ref{ss:sm_obs} into Eq.~(\ref{eq:decoup_2}), we obtain the boundary condition
\begin{equation}
    \label{eq:sm_decoup_2b}
    \Phi'(0^\pm) \pm \frac{2 \hat{\psi}}{\e^{\pm \lambda} - 1} \Phi(0^\pm) = 0
    \:,
\end{equation}
which is completed by a second boundary condition at infinity,
\begin{equation}
    \label{eq:sm_decoup_4b}
    \Phi(v) \xrightarrow[v\to\pm\infty]{} \rho
    \:.
\end{equation} }
We stress that these equations are exact in the hydrodynamic limit considered here.

\subsection{Known results {on the GDP generating function}}

In this Section, we focus on the bulk equation~\eqref{eq:sm_decoup_1b}. We recall here the results of~\cite{poncet2021generalisedSI} both in the high and low density regimes, in which this equation simplifies. This will be the starting point to tackle the arbitrary density case.

\subsubsection{High density}

In the high density limit $\rho\to 1$, we define
\begin{equation}
    \check\Phi(v) = \lim_{\rho\to 1} \frac{\Phi(v)}{1-\rho}.
\end{equation}
In this limit, the higher order correlations $C(v)$ can be neglected, and the bulk equation~\eqref{eq:sm_decoup_1b} becomes~\cite{poncet2021generalisedSI}
\begin{equation}
\label{eq:bulkHighDens}
    \check \Phi''(v) + 2v \check \Phi'(v) = 0.
\end{equation}

\subsubsection{Low density}

In the opposite limit of low density, $\rho\to 0$, one should keep $z = \rho r$ and $\tau = \rho^2 t$ constant. With these scalings, we define
\begin{align}
    \label{eq:ScalingsPhiLowDens}
	\hat \Phi(v, \hat\lambda) &= \lim_{\rho\to 0} \frac{\Phi(v, \lambda = \rho\hat\lambda)}{\rho}
	\:, 
	&
	\beta(\hat\lambda) &= \lim_{\rho\to 0} \frac{\pm b_\pm(\lambda =\rho\hat\lambda)}{\rho}
	= \lim_{\rho\to 0} \frac{\hat{\psi}(\rho \hat{\lambda})}{\rho \hat{\lambda}}
	\:.
\end{align}
In this limit, the bulk equation~\eqref{eq:sm_decoup_1b} is not closed. In~\cite{poncet2021generalisedSI}, a closure relation was found, which gives
\begin{equation} \label{eq:decoup_1_low_c}
\hat \Phi''(v) + 2(v+\xi) \hat\Phi'(v) = 0,
\end{equation}
with $\xi$ the (rescaled) derivative of the cumulant-generating function with respect to its parameter,
\begin{equation}
\label{eq:xiLowDens}
\xi \equiv \beta + \hat\lambda \frac{\dd\beta}{\dd\hat\lambda} 
= \frac{\dd}{\dd\hat\lambda}(\hat\lambda \beta)
\:.
\end{equation}

\subsubsection{Lowest order}

At order $1$ in $\lambda$, the equation~\eqref{eq:sm_decoup_1b} is also closed~\cite{poncet2021generalisedSI}
\begin{equation}
    \label{eq:sm_EqOrder1}
    \Phi_1''(v) + 2v \Phi_1'(v) = 0,
\end{equation}
and its solution gives
\begin{equation}
\label{eq:sm_Phi1_SEP}
    \Phi_1(v\gtrless 0)  = \pm \frac{1-\rho}{2} \erfc |v|.
\end{equation}

\section{SEP at arbitrary density}

\subsection{A closed integral equation}

{
The bulk equation~\eqref{eq:sm_decoup_1b}, valid at arbitrary density, is not closed: in addition to the functions $\hat{\psi}$ and $\Phi(v)$ of interest, it involves the unknown function $C(v)$. We thus look for a closed equation for $\Phi$, aiming}
to extend the bulk equations obtained at high density~\eqref{eq:bulkHighDens} and low density~\eqref{eq:decoup_1_low_c} to arbitrary density. { More precisely, we guess that this closed equation is of the form}
\begin{equation}
\label{eq:sm_expectedEqPhi}
  \Phi''(v) + 2(v+\xi) \Phi'(v) =
  \circled{?}
  \:,
\end{equation}
with a right hand side to be determined, and which vanishes in both limits $\rho \to 0$ and $\rho \to 1$. For this equation to be closed, $\circled{?}$ should be expressed in terms of $\Phi(v)$ and the parameters $\lambda$ and $\xi$ only. Furthermore, we should also have $\circled{?} = 0$ at order $1$ in $\lambda$, because of~\eqref{eq:sm_EqOrder1}. We thus expect that, at order $n$, the r.h.s. $\circled{?}$ will act as a source term for the determination of $\Phi_n$, by involving only the profiles $\Phi_m$ with $m < n$. { Furthermore}, the resulting equation, combined with the boundary conditions~(\ref{eq:sm_decoup_3b},\ref{eq:sm_decoup_2b},\ref{eq:sm_decoup_4b}), should also reproduce the cumulants of the tracer's position obtained recently in~\cite{Imamura:2017SI,Imamura:2021SI}. An interesting feature of these cumulants is that they involve nontrivial factors $\sqrt{2}$ (for the fourth cumulant $\kappa_4$) and $\sqrt{3}$ (for $\kappa_6$), which cannot be produced by products or derivatives of $\Phi_1$~\eqref{eq:sm_Phi1_SEP}. These factors can however be obtained by considering half-convolutions of $\Phi_1$ with itself, such as
\begin{equation}
    \int_0^\infty \dd z \: \Phi_1'(v+z) \Phi_1'(-z)
    = \frac{(1-\rho)^2}{\sqrt{2\pi}} \e^{-\frac{v^2}{2}} \erfc \left( \frac{v}{\sqrt{2}} \right)
    \:.
\end{equation}
{We also have some constraints on how the different parameters ($\lambda$, $\hat{\psi}$ and $\xi$) should appear in the desired equation. For instance, $\lambda$ is explicitly involved in the hydrodynamic equations of Section~\ref{ss:sm_eqs2} only through expressions of the form $\e^{\pm \lambda}-1$, so we expect that only these kind of expressions appear. In the low density equation~\eqref{eq:decoup_1_low_c}, $\hat{\psi}$ does not appear explicitly, only its derivative $\xi = \frac{\dd \hat{\psi}}{\dd \lambda}$ is involved, so we expect the same to happen at arbitrary density.

Finally, the equation we write should have a "proper scaling" with time. Indeed, the bulk equation~\eqref{eq:sm_decoup_1b} (which we aim to replace) is obtained by expanding at order $1/t$ the microscopic equation~\eqref{eq:sm_decoup_1}, so it should be the same for this new equation. For instance, the functions $\Phi$, $\Phi'$ and $\Phi''$ are respectively of orders $t^0$, $t^{-1/2}$ and  and $t^{-1}$. The scaling argument $v = r/\sqrt{2t}$ is of order $t^{-1/2}$, and the same scaling holds for $\xi$. One can thus check that, with these scalings, the l.h.s. of Eq.~\eqref{eq:sm_expectedEqPhi} indeed has the correct scaling $1/t$. The same should hold for the r.h.s. $\circled{?}$.

Following these ideas and constraints}, we obtained that the equation (valid for $v>0$)
\begin{multline}
  \label{eq:EqPhiOrder6p}
   \Phi''(v) + 2(v+\xi) \Phi'(v)
  =
   \frac{\e^{-\lambda}-1}{\Phi(0^-)}
    \int_{0}^\infty \dd z \:
    \Phi''(-z) \Phi'(v+z)
  +
  2\xi \frac{\e^{-\lambda}-1}{\Phi(0^-)}
  \int_{0}^\infty \dd z \:
  \Phi'(-z) \Phi'(v+z)
  \\
  + \frac{(\e^{\lambda}-1)(\e^{-\lambda}-1)}{\Phi(0^+)\Phi(0^-)}
  \int_0^\infty \dd z \int_0^\infty \dd z'
  \: \Phi''(-z-z') \Phi'(z+v) \Phi'(z')
  \\
  - \frac{(\e^{-\lambda}-1)^2}{\Phi(0^-)^2} \int_0^\infty \dd z \int_0^\infty \dd z'
  \: \Phi''(z+z'+v) \Phi'(-z) \Phi'(-z')
  \\
  + 2 \xi \frac{(\e^{\lambda}-1)(\e^{-\lambda}-1)}{\Phi(0^+)\Phi(0^-)}
  \int_0^\infty \dd z \int_0^\infty \dd z'
  \: \Phi'(-z-z') \Phi'(z+v) \Phi'(z')
  \\
  - 2 \xi \frac{(\e^{-\lambda}-1)^2}{\Phi(0^-)^2} \int_0^\infty \dd z \int_0^\infty \dd z'
  \: \Phi'(z+z'+v) \Phi'(-z) \Phi'(-z')
  + \cdots,
\end{multline}
properly reproduces the known cumulant $\kappa_n$ for $n \leq 6$. The equation for $v<0$ is deduced from the symmetry $\Phi(-v,\lambda) = \Phi(v,-\lambda)$. This leads us to introduce the two functions
\begin{subequations}
  \label{eq:defNewFct}
  \begin{align}
    \Omega_+(v)
    &= \frac{\e^{\lambda}-1}{\Phi(0^+)} \Phi'(v)
    = - 2 \hat{\psi} \frac{\Phi'(v)}{\Phi'(0^+)}
     \quad \text{ for } v > 0 \:,
    \\
    \Omega_-(v)
    &= \frac{\e^{-\lambda}-1}{\Phi(0^-)} \Phi'(v)
    = 2 \hat{\psi} \frac{\Phi'(v)}{\Phi'(0^-)}
     \quad \text{ for } v < 0 \:,
  \end{align}  
\end{subequations}
so that~\eqref{eq:EqPhiOrder6p} takes the more compact form
\begin{multline}
  \label{eq:EqPhiOrder6pNew}
  \Omega_+'(v) + 2(v+\xi) \Omega_+(v)
  =
  \int_{0}^\infty \dd z \:
  \Omega_-'(-z) \Omega_+(v+z)
  +
  2\xi \int_{0}^\infty \dd z \:
  \Omega_-(-z) \Omega_+(v+z)
  \\
  + \int_0^\infty \dd z \int_0^\infty \dd z'
  \: \Omega_-'(-z-z') \Omega_+(z+v) \Omega_+(z')
  + 2 \xi
  \int_0^\infty \dd z \int_0^\infty \dd z'
  \: \Omega_-(-z-z') \Omega_+(z+v) \Omega_+(z')
  \\
  - \int_0^\infty \dd z \int_0^\infty \dd z'
  \: \Omega_+'(z+z'+v) \Omega_-(-z) \Omega_-(-z')
  - 2 \xi \int_0^\infty \dd z \int_0^\infty \dd z'
  \: \Omega_+(z+z'+v) \Omega_-(-z) \Omega_-(-z')
  +\cdots
\end{multline}
We can rewrite this expression in terms of the matrix operator $\mathscr{L}$ defined as
\begin{equation}
  \label{eq:DefOpL}
  \mathscr{L} =
  \begin{pmatrix}
    \mathscr{L}_{++} & \mathscr{L}_{+-} \\
    \mathscr{L}_{-+} & \mathscr{L}_{--}
  \end{pmatrix}
  \:,
\end{equation}
with
\begin{subequations}
  \label{eq:DefOpLOmega}
  \begin{align}
    (\mathscr{L}_{++}f)(v)
    &=
       \int_{0}^\infty \dd z \: \Omega_-(-z) f(v+z)
      \:,
    &
    (\mathscr{L}_{+-}f)(v)
    &=
      \int_{0}^\infty \dd z \: \Omega_+(v+z) f(-z) 
      \:,
    \\
    (\mathscr{L}_{-+}f)(v)
    &=
     -\int_{0}^\infty \dd z \: \Omega_-(v-z) f(z)
      \:,
    &
    (\mathscr{L}_{--}f)(v)
    &=
      -\int_{0}^\infty \dd z \: \Omega_+(z) f(v-z)
      \:.
  \end{align}
\end{subequations}
Applying this operator to the column vector $( \Omega_+ \: 0)^{\mathrm{T}}$, we get
\begin{equation}
  \mathscr{L}
  \begin{pmatrix}
    \Omega_+ \\ 0
  \end{pmatrix}
  =
  \begin{pmatrix}
    \displaystyle
     \int_{0}^\infty \dd z \: \Omega_-(-z) \Omega_+(v+z)
     \\[0.3cm]
    \displaystyle
     -\int_{0}^\infty \dd z \: \Omega_-(v-z)\Omega_+(z)
  \end{pmatrix}
  \:,
\end{equation}
whose first component appears in our equation~\eqref{eq:EqPhiOrder6pNew}. Applying $\mathscr{L}^2$ to the same vector, we get for the first component,
\begin{equation}
  \int_0^\infty \dd z \int_0^\infty \dd z' \: \Omega_+(z+z'+v) \Omega_-(-z) \Omega_-(-z')
  - \int_0^\infty \dd z \int_0^\infty \dd z' \: \Omega_-(-z-z') \Omega_+(z+v) \Omega_+(z')
  \:,
\end{equation}
which corresponds to the next terms in Eq.~\eqref{eq:EqPhiOrder6pNew}.
Finally, after some integration by parts, and using that
\begin{equation}
  \Omega_+(0) = - \Omega_-(0)
  \:,
\end{equation}
we can write~(\ref{eq:EqPhiOrder6pNew}) as
\begin{equation}
  \label{eq:ClosedEqOmegP}
  \boxed{
    2 v \Omega_+(v) +
    (\partial_v + 2\xi) \left[ (1+\mathscr{L})^{-1}
      \begin{pmatrix}
        \Omega_+ \\ 0
      \end{pmatrix} (v)
    \right]_1
    + \Omega_+(v)  \left[ (1+\mathscr{L})^{-1}
      \begin{pmatrix}
        \Omega_+ \\ 0
      \end{pmatrix} (0)
    \right]_1
    = 0
    \:.
  }
\end{equation}
Similarly, the equation for $v<0$ takes the form
\begin{equation}
  \label{eq:ClosedEqOmegM}
  \boxed{
    2 v \Omega_-(v) +
    (\partial_v + 2\xi) \left[ (1+\mathscr{L})^{-1}
      \begin{pmatrix}
        0 \\ \Omega_-
      \end{pmatrix} (v)
    \right]_2
    + \Omega_-(v)  \left[ (1+\mathscr{L})^{-1}
       \begin{pmatrix}
        0 \\ \Omega_-
      \end{pmatrix} (0)
    \right]_2
    = 0
    \:.
  }
\end{equation}
In order to simplify these equations, we introduce another set of functions $M_{\pm,\pm} $ defined by
\begin{equation}
  \label{eq:DefM}
  (1+\mathscr{L})^{-1}
  \begin{pmatrix}
    \Omega_+ & 0 \\
    0 & \Omega_-
  \end{pmatrix}
  =
  \begin{pmatrix}
    M_{++} & M_{+-} \\
    M_{-+} & M_{--}
  \end{pmatrix}
  \:.
\end{equation}
Expanding the above expression in powers of $\mathscr{L}$, we notice that
\begin{equation}
  \label{eq:RelMOmega}
  \Omega_+(v) = M_{++}(v) - M_{+-}(v)
  \:,
  \quad
  \Omega_-(v) = M_{--}(v) - M_{-+}(v)
  \:.
\end{equation}
Using these results, we can rewrite the equations~(\ref{eq:ClosedEqOmegP},\ref{eq:ClosedEqOmegM}) as
\begin{equation}
  \label{eq:EqMpp}
  (\partial_v + 2(v+\xi) + M_{++}(0)) M_{++} = (2 v + M_{++}(0)) M_{+-}
  \:,
\end{equation}
\begin{equation}
  \label{eq:EqMmm}
  (\partial_v + 2(v+\xi) + M_{--}(0)) M_{--} = (2 v + M_{--}(0)) M_{-+}
  \:.
\end{equation}
We can obtain a closed system of equations for $M$ by rewriting~(\ref{eq:DefM}) as
\begin{equation}
  \label{eq:ActLonM}
  \mathscr{L}
  \begin{pmatrix}
    M_{++} & M_{+-} \\
    M_{-+} & M_{--}
  \end{pmatrix}
  + \begin{pmatrix}
    M_{+-} & M_{+-} \\
    M_{-+} & M_{-+}
  \end{pmatrix}
  = 0
  \:,
\end{equation}
where the operator $\mathscr{L}$ can be written in terms of $M$ only
by using~(\ref{eq:RelMOmega}). The two columns are identical,
therefore this only gives two equations, which are
\begin{equation}
  \label{eq:IntegEqMpm}
  M_{+-}(v) = - \int_0^\infty \dd z \left[
    M_{++}(v+z) M_{--}(-z) - M_{+-}(v+z) M_{-+}(-z)
  \right]
  \:,
  \quad \text{for } v > 0
  \:,
\end{equation}
\begin{equation}
  \label{eq:IntegEqMmp}
  M_{-+}(v) = \int_0^\infty \dd z \left[
    M_{--}(v-z) M_{++}(z) - M_{+-}(z) M_{-+}(v-z)
  \right]
  \:,
  \quad \text{for } v < 0
  \:.
\end{equation}
In order to proceed further, it is instructive to look for a perturbative solution of these equations. By definition~\eqref{eq:defNewFct}, $\Omega_\pm$ (and thus $M_{\pm,\pm}$) is small when $\hat{\psi}$ is small. We find that the solutions of~(\ref{eq:EqMpp},\ref{eq:EqMmm},\ref{eq:IntegEqMpm},\ref{eq:IntegEqMmp}) at first orders in $\lambda$ can be conveniently expressed as
\begin{subequations}
  \begin{align}
    M_{++}(v)
    &=
      \left( -\omega \: \e^{\xi^2} \right) \Omega_+^{(1)}(v)
      +2\left( -\omega \: \e^{\xi^2} \right)^2 \Omega_+^{(2)}(v)
      +3\left( -\omega \: \e^{\xi^2} \right)^3 \Omega_+^{(3)}(v)
      + \O(\omega^4)
      \:,
    \\
    M_{+-}(v)
    &=
      \left( -\omega \: \e^{\xi^2} \right)^2 \Omega_+^{(2)}(v)
      +2\left( -\omega \: \e^{\xi^2} \right)^3 \Omega_+^{(3)}(v)
      + \O(\omega^4)
      \:,
  \end{align}
\end{subequations}
where we introduced the parameter $\omega$ defined from $\hat{\psi}$ as
\begin{equation}
\label{eq:expXpsi}
  \omega =  2 \hat{\psi} + \sqrt{2\pi} \: \hat{\psi}^2 +
  \frac{2\pi}{9} (9 - 4 \sqrt{3}) \hat{\psi}^3 + \O(\hat{\psi}^4)
  \quad \Rightarrow \quad
  \hat{\psi} = \frac{\omega}{2}
  - \frac{\sqrt{\pi}}{4 \sqrt{2}} \omega^2
  + \frac{\pi}{6 \sqrt{3}} \omega^3 + \O(\omega^4)
  \:
\end{equation}
and
\begin{subequations}
\label{eq:sm_FirstOrdersOmegaX}
  \begin{align}
    \label{eq:ExprOmPOrder1X}
    \Omega_\pm^{(1)}(v)
    &= \pm \e^{-(v+\xi)^2}
      \:,
    \\
     \Omega_\pm^{(2)}(v)
    &= \pm \frac{1}{2} \sqrt{\frac{\pi}{2}}
      \e^{-\frac{1}{2}(v+2\xi)^2} \erfc \left( \pm\frac{v}{\sqrt{2}} \right)
      \:,
    \\
    \Omega_\pm^{(3)}(v)
    &= \pm \frac{\pi}{4 \sqrt{3}} \e^{- \frac{1}{3}(v+3\xi)^2}
      \left[
      \erfc\left( \pm\sqrt{\frac{2}{3}} v \right)
      + \erfc\left( \pm\frac{v}{\sqrt{6}} \right)
      - 4 \mathrm{T} \left( \frac{v}{\sqrt{3}}, \sqrt{3} \right)
      \right]
      \:,
  \end{align}
\end{subequations}
with $\mathrm{T}$ the Owen's T-function defined as~\cite{Owen:1980SI}
\begin{equation}
  \label{eq:defOwenT}
  \mathrm{T}(h,a) = \frac{1}{2\pi}
  \int_0^a \frac{\e^{-\frac{h^2}{2}(1+x^2)}}{1+x^2} \dd x
  \:.
\end{equation}
This leads us to write the general form as
\begin{equation}
  \label{eq:GenFormExpM}
  M_{++}(v) = \sum_{n \geq 1} n \left( -\omega \: \e^{\xi^2} \right)^n \Omega_+^{(n)}(v)
  \:,
  \quad
  M_{+-}(v) = \sum_{n \geq 2} (n-1) \left( -\omega \: \e^{\xi^2} \right)^n \Omega_+^{(n)}(v)
  \:,
\end{equation}
so that, from~\eqref{eq:RelMOmega},
\begin{equation}
  \label{eq:GenFormExpOmega}
  \Omega_+(v) = \sum_{n \geq 1}  \left( -\omega \: \e^{\xi^2} \right)^n \Omega_+^{(n)}(v)
  \:,
\end{equation}
and similarly for $\Omega_-$. Plugging these expressions into~\eqref{eq:IntegEqMpm}, we obtain
\begin{equation}
  \Omega_+^{(n)}(v)
  \\
  = - \sum_{p=1}^{n-1}
  \int_{0}^\infty \dd z \:
  \Omega_+^{(p)}(v+z) \Omega_-^{(n-p)}(-z)
  \:.
\end{equation}
Multiplying by $(-\omega \e^{\xi^2})^n$ and summing over $n$, we get
\begin{equation}
  \Omega_+(v) - 
  \left( -\omega \: \e^{\xi^2} \right) \Omega_+^{(1)}(v)
  = - \int_0^\infty \dd z \: \Omega_+(v+z) \Omega_-(-z)
  \:.
\end{equation}
From the expression of $\Omega_+^{(1)}$~\eqref{eq:ExprOmPOrder1X}, this becomes
\begin{equation}
  \label{eq:sm_IntegEqOmP}
    \Omega_+(v) + \omega \: \e^{-(v+\xi)^2 + \xi^2}
    + \int_0^\infty \dd z \: \Omega_+(v+z) \Omega_-(-z)
    = 0
    \:,
\end{equation}
and similarly we obtain the equation for $\Omega_-$ from~(\ref{eq:IntegEqMmp}):
\begin{equation}
  \label{eq:sm_IntegEqOmM}
    \Omega_-(v) - \omega \: \e^{-(v+\xi)^2 + \xi^2}
    - \int_0^\infty \dd z \: \Omega_-(v-z) \Omega_+(z)
    = 0
    \:.
\end{equation}
Equations similar to~(\ref{eq:sm_IntegEqOmP},\ref{eq:sm_IntegEqOmM}) can be found in Ref.~\cite{Arabadzhyan:1987SI}. In this paper, the authors show that these two coupled equations are equivalent to the two independent linear equations
\begin{equation}
  \label{eq:sm_IntegEqOmPOnly}
  \boxed{
    \Omega_+(v) = -\omega \: \e^{-(v+\xi)^2 + \xi^2}
    - \omega \int_{-\infty}^0 \dd z \: \Omega_+(z) \: \e^{-(v-z+\xi)^2 + \xi^2}
    \:,
  }
\end{equation}
\begin{equation}
  \label{eq:sm_IntegEqOmMOnly}
  \boxed{
    \Omega_-(v) = \omega \: \e^{-(v+\xi)^2 + \xi^2}
    -\omega  \int_0^{\infty} \dd z \: \Omega_-(z) \: \e^{-(v-z+\xi)^2 + \xi^2}
    \:,
  }
\end{equation}
which correspond to the Eq.~(6) 
given in the main text. These equations give $\Omega_\pm$ for all $v \in \mathbb{R}$ (including the analytic continuations of $\Omega_+$ to $v<0$ and $\Omega_-$ to $v>0$ which both appear explicitly in the integrals). 

\subsection{Solution}

Equations similar to~(\ref{eq:sm_IntegEqOmPOnly},\ref{eq:sm_IntegEqOmMOnly}) are solved in~\cite{Polyanin:2008SI}, but restricted to $v>0$, and with $\Omega_\pm(\mp v) = 0$ for $v<0$. We can nevertheless use these results to express $\Omega_\pm$ in this domain. The results are given in terms of the one-sided Fourier transforms:
\begin{equation}
  \hat{\Omega}_\pm^{(+)}(k) =
  \int_0^\infty \dd v \: \Omega_\pm(v) \: \e^{\I k v}
  \:,
  \quad
  \hat{\Omega}_\pm^{(-)}(k) =
  -\int_{-\infty}^0 \dd v \: \Omega_\pm(v) \: \e^{\I k v}
  \:.
\end{equation}
From Ref.~\cite{Polyanin:2008SI}, we obtain the Fourier transforms of the
analytic continuations of $\Omega_\pm$:
\begin{subequations}
  \begin{align}
    \hat{\Omega}_+^{(-)}(k)
    &= 1 - \exp \left[
      \frac{1}{2} \sum_{n \geq 1}
      \frac{(-\omega \sqrt{\pi} \: \e^{-\frac{1}{4}(k+2\I \xi)^2})^n}{n}
      \erfc \left(- \sqrt{n} \left(\xi - \frac{\I k}{2} \right) \right)
      \right]
      \:,
    \\
    \hat{\Omega}_-^{(+)}(k)
    &= 1 - \exp \left[
      \frac{1}{2} \sum_{n \geq 1}
      \frac{(-\omega \sqrt{\pi} \: \e^{-\frac{1}{4}(k+2\I \xi)^2})^n}{n}
      \erfc \left(\sqrt{n} \left(\xi - \frac{\I k}{2} \right) \right)
      \right]
      \:.
  \end{align}
\end{subequations}
We can obtain the Fourier transforms on the original functions by
taking the Fourier transform of~(\ref{eq:sm_IntegEqOmPOnly}), which gives
\begin{equation}
  \hat{\Omega}_+^{(+)}(k) 
  = \hat{K}(k) + \hat{\Omega}_+^{(-)}(k) ( 1 - \hat{K}(k))
  \:,
  \quad \text{with} \quad
  \hat{K}(k) = - \omega \sqrt{\pi} \: \e^{- \frac{1}{4}(k+2\I \xi)^2}
  \:.
\end{equation}
We finally obtain
\begin{equation}
  \label{eq:sm_SolOmPFourier}
  \boxed{
    \int_0^\infty  \Omega_+(v) \e^{\I k v} \dd v
    = 1 - \exp \left[- Z_+ \left( \omega, \xi - \frac{\I k}{2} \right) \right]
    \:,
  }
\end{equation}
\begin{equation}
  \label{eq:sm_SolOmMFourier}
  \boxed{
    \int_{-\infty}^0  \Omega_-(v) \e^{\I k v} \dd v
    = \exp \left[- Z_- \left( \omega, \xi - \frac{\I k}{2} \right) \right] - 1
    \:,
  }
\end{equation}
where we have introduced
\begin{equation}
\label{eq:sm_defZpm}
    Z_\pm(\omega,\xi) = \frac{1}{2} \sum_{n \geq 1}
      \frac{(-\omega \sqrt{\pi} \: \e^{\xi^2})^n}{n}
      \erfc \left( \pm \sqrt{n} \xi  \right)
      \:.
\end{equation}
We can obtain the values of $\Omega_\pm(0) = \mp 2 \hat{\psi}$   from these expressions. We set $k = \I s$ in~(\ref{eq:sm_IntegEqOmPOnly}) and let $s \to \infty$, this gives
\begin{equation}
\label{eq:sm_relPsiomegaPolyLog}
  \Omega_+(0) = - 2 \hat{\psi} =
  \frac{1}{\sqrt{\pi}} \mathrm{Li}_{\frac{3}{2}} (- \omega \sqrt{\pi})
  \:,
\end{equation}
which expresses $\hat{\psi}$ in terms of $\omega$. Note that this expression is consistent with the first orders obtained previously~\eqref{eq:expXpsi}.

\subsection{Expansion in \texorpdfstring{$\lambda$}{lambda}}
\label{sec:ExpLambdaSEP}

\subsubsection{Computation of the cumulants}

Plugging the expansion of $\hat{\psi}$~\eqref{eq:DefPsiAndExpLamb} into~\eqref{eq:sm_relPsiomegaPolyLog}, we can deduce the expansion of $\omega$ in powers of $\lambda$, in terms of the cumulants $\kappa_n$. Combining the solution~\eqref{eq:sm_SolOmPFourier} (for $k=0$) with the definition of $\Omega_+$~\eqref{eq:defNewFct}, and the boundary condition~\eqref{eq:sm_decoup_2b}, we obtain
\begin{equation}
\label{eq:sm_RelIntOmPhi0}
    1 - \exp \left[
      -\frac{1}{2} \sum_{n \geq 1}
      \frac{(-\omega \sqrt{\pi} \: \e^{\xi^2})^n}{n}
      \erfc \left(\sqrt{n}\xi  \right)
    \right]
    = (\e^\lambda-1) \frac{\rho - \Phi(0^+)}{\Phi(0^+)}
    \:.
\end{equation}
Since $\xi = \frac{\dd \hat{\psi}}{\dd \lambda}$, this equation gives $\Phi_n(0^+)$ in terms of the cumulants $\kappa_j$ for $j \leq n$. We can proceed similarly for $\Phi_n(0^-)$ using~\eqref{eq:sm_SolOmMFourier}. Finally, using the last relation~\eqref{eq:sm_decoup_3b}, we can determine the cumulants $\kappa_n$ and thus $\Phi_n(0^\pm)$. Due to the symmetry $v\to -v$ with $\lambda \to -\lambda$, all the odd order cumulants vanish. For the even order ones, we get for instance,
\begin{equation}
  \kappa_2 = \frac{1-\rho}{\rho \sqrt{\pi}}
  \:,
\end{equation}
\begin{equation}
  \kappa_4 = \frac{(1-\rho)}{\pi^{3/2} \rho^3}
  \left(
    12 (1-\rho)^2 - \pi (3 - 3 (4 - \sqrt{2})\rho + (8-3\sqrt{2}) \rho^2)
  \right)
  \:,
\end{equation}
\begin{multline}
  \kappa_6 = 
  \frac{(1-\rho )}{\pi^{5/2} \rho^5}
  \left(
    30 \pi  \left(2 \left(9 \sqrt{2}-20\right) \rho^2+\left(60-18 \sqrt{2}\right) \rho -15\right) (1-\rho )^2
  \right.
  \\
  -\pi^2 \left(
    8 \left(-17+15 \sqrt{2}-5 \sqrt{3}\right) \rho^4
    +\left(480-300 \sqrt{2}+80 \sqrt{3}\right) \rho ^3
  \right.
  \\
  \left.
    \left.
      +5 \left(-114+45 \sqrt{2}-8 \sqrt{3}\right) \rho^2
      -45 \left(\sqrt{2}-6\right) \rho -45\right)+1020 (1-\rho )^4
  \right)
  \:,
\end{multline}
which coincide with the cumulants obtained from the CGF computed in~\cite{Imamura:2017SI}. This is expected because the equation~\eqref{eq:EqPhiOrder6p} has been constructed in order to reproduce these cumulants. We have further checked with Mathematica, up to $n=10$, that the next cumulants $\kappa_n$ obtained by our procedure also coincide with those obtained from~\cite{Imamura:2017SI}. This provides a nontrivial validation of our integral equations~(\ref{eq:sm_IntegEqOmPOnly},\ref{eq:sm_IntegEqOmMOnly}).

\subsubsection{Computation of the generalized profiles}

Having determined $\Phi_n(0^\pm)$ and $\kappa_j$ for $j \leq n$, and thus $\Phi_n'(0^\pm)$ from the boundary condition~\eqref{eq:sm_decoup_2b}, we can express $\Phi'(v)$ in terms of $\Omega_\pm(v)$. Expanding the explicit solutions~(\ref{eq:sm_SolOmPFourier},\ref{eq:sm_SolOmMFourier}) in powers of $\omega$ and computing the inverse Fourier transform, we recover the expansion~\eqref{eq:GenFormExpOmega} with $\Omega_\pm^{(n)}(v)$ given by~\eqref{eq:sm_FirstOrdersOmegaX} for $n \leq 3$, but we can also access higher orders via an inverse Fourier transform. Having expressed $\omega$ in terms of the cumulants determined previously via~\eqref{eq:sm_relPsiomegaPolyLog}, we thus have the expansion of $\Phi'(v)$ in powers of $\lambda$. After integration, we obtain in particular
\begin{subequations}
  \label{eq:PhiFirstOrders}
  \begin{align}
    \label{eq:Phi0}
  \Phi_0(v)
  &= \rho \:,
  \\
    \label{eq:Phi1}
  \Phi_1(v)
  &= \frac{1-\rho}{2} \erfc(v) \:,
  \\
    \label{eq:Phi2}
  \Phi_2(v)
  &= \frac{(1-\rho)(1-2\rho)}{2\rho} \erfc(v)
    - \frac{2}{\pi} \frac{(1-\rho)^2}{\rho} \: \e^{-v^2}
    \:,
\end{align}
\end{subequations}
\begin{multline}
  \label{eq:Phi3}
  \Phi_3(v)
  =
  (1-\rho )\frac{2 (3+\pi ) \rho ^2-(12+\pi ) \rho +6 + \pi \rho(1-\rho)}
  {2 \pi  \rho ^2}
  \erfc(v)
  \\
  +3 (1-\rho)^2  \frac{2 (1-\rho) v - \sqrt{\pi } (1-2 \rho)}
  {\pi^{3/2} \rho ^2} \e^{-v^2}
  -\frac{3(1-\rho)^2}{4\rho} \erfc \left( \frac{v}{\sqrt{2}} \right)^2
  \:,
\end{multline}
\begin{multline}
  \label{eq:Phi4}
  \Phi_4(v) =(1-\rho)(1-2\rho) \frac{24 (1-\rho)^2 - \pi (3 (1-2\rho)^2 + 4
    (1-\rho)(1-2\rho)+18\rho (1-\rho)-4)}{2\pi \rho^3} \erfc(v)
  \\
  + 4(1-\rho)^2 \frac{3 \sqrt{\pi} (1-\rho)(1-2\rho)v
    -4(1-\rho)^2 (4+v^2) + \pi(5 (1-\rho)(1-2\rho) + 3 (1+\sqrt{2}) \rho(1-\rho)-2)
  }{\pi^2 \rho^3}\e^{-v^2}
  \\
  + 12 \sqrt{2} \frac{(1-\rho)^3}{\pi \rho^2} \e^{-\frac{v^2}{2}}
  \erfc \left( \frac{v}{\sqrt{2}} \right)
  - 3 \frac{(1-\rho)^2(1-2\rho)}{2 \rho^2}  \erfc \left( \frac{v}{\sqrt{2}} \right)^2
  \:.
\end{multline}

\begin{multline}
    \Phi_5(v) = -\frac{(1-\rho) \left(\pi ^2 \rho ^4
    +440 (1-\rho)^4+30 \pi  \left(2 \rho  \left(\left(2 \sqrt{2}-3\right) \rho-2\sqrt{2}+6\right)-3\right) 
    (1-\rho)^2\right)}{240 \pi^2 \rho^4}
   \erfc(v)
   \\
   + \frac{(1-\rho)^2}{24 \pi ^{5/2} \rho ^4} \left[
      4 \sqrt{\pi } (1-\rho)^2 \left(19 \rho +(4 \rho -2) v^2-11\right)
      +12 \pi  \left(\rho \left(\left(2 \sqrt{2}-5\right) \rho -2 \sqrt{2}+8\right)-2\right) (1-\rho) v
   \right.
   \\
   \left.
   -3 \pi ^{3/2} (1-2 \rho) \left(2 \rho \left(\left(\sqrt{2}-3\right) \rho -\sqrt{2}+5\right)-3\right)
   +4 (1-\rho)^3 v \left(2 v^2+27\right)
   \right]
   \e^{-v^2}
   \\
   + \frac{(1-\rho)^3\left(\sqrt{\pi } (1-2 \rho)-2 (1-\rho) v\right)}{2 \sqrt{2} \pi ^{3/2} \rho ^3}
   \e^{-\frac{v^2}{2}}
   \erfc \left(\frac{v}{\sqrt{2}}\right) 
   + \frac{(1-\rho)^2 \left(\pi  \left(2 \rho ^2-4 \rho +1\right)-12 (1-\rho)^2\right)}{32 \pi \rho ^3}
   \erfc\left(\frac{v}{\sqrt{2}}\right)^2
   \\
   - \frac{(1-\rho)^3}{4 \sqrt{3 \pi} \: \rho^2} 
   \int_v^{+\infty} \dd z \: \e^{-\frac{z^2}{3}} 
    \left[
      \erfc\left( \sqrt{\frac{2}{3}} z \right)
      + \erfc\left( \frac{z}{\sqrt{6}} \right)
      - 4 \mathrm{T} \left( \frac{z}{\sqrt{3}}, \sqrt{3} \right)
      \right]
   \:,
\end{multline}
with the Owen-T function defined in~\eqref{eq:defOwenT}.

\subsubsection{Conservation relation}

Using the results above, the conservation relation
\begin{equation}
\label{eq:sm_ConsRel}
    \int_0^\infty (\Phi(v) - \rho) \dd v 
    - \int_{-\infty}^0 (\Phi(v) - \rho) \dd v
    = \rho \xi
\end{equation}
holds up to $\O(\lambda^6)$, and non perturbatively in $\lambda$ (numerically).

\subsection{Extensions}

\subsubsection{Step density profile}

Our formalism can be extended to the case of an initial step density $\rho_-$ for $v<0$ and $\rho_+$ for $v>0$, by only changing the boundary condition at infinity~\eqref{eq:sm_decoup_4b} into $\Phi(\pm \infty) = \rho_\pm$. Unlike the constant density case, the odd order cumulants do not vanish here. We can still apply the procedure described in Section~\ref{sec:ExpLambdaSEP}, which gives that $\kappa_1$ is solution of
\begin{equation}
    \frac{\rho_-}{1 + \sqrt{\pi} \: \kappa_1 \e^{\kappa_1^2} \erfc(-\kappa_1)}
    =
    \frac{\rho_+}{1 - \sqrt{\pi} \: \kappa_1 \e^{\kappa_1^2} \erfc(\kappa_1)}
    \:,
\end{equation}
and the higher order cumulants are expressed in terms of $\kappa_1$. For instance,
\begin{equation}
\kappa_2 = 
\kappa _1^2 \left(2 \pi  \e^{2 \kappa_1^2} \kappa_1 \erfc\left(\sqrt{2} \kappa_1\right)
+\frac{4 \pi  \e^{2 \kappa_1^2} \kappa_1 \rho_+ \left(\rho_+^2-3 \rho_- \rho_+ +2 \rho_-\right)}
{\left(\rho_- -\rho_+\right)^3}
-\sqrt{2 \pi }\right)
    \:,
\end{equation}
These expressions coincide with those given in~\cite{Imamura:2017SI}. We additionally obtain the profiles $\Phi_n(v)$, for instance
\begin{equation}
    \Phi_0(v) = \frac{\rho_+}{2} \erfc(-v-\kappa_1) + \frac{\rho_-}{2} \erfc(v+\kappa_1)
    \:,
\end{equation}
\begin{multline}
    \Phi_1(v) = \frac{2 \sqrt{\pi } \e^{\kappa _1^2} \kappa _1 \rho _-
   \left(1-\rho_+\right)-\left(\rho _-\rho _+\right)^2}{2 \left(\rho _--\rho_+\right)}
   \erfc \left( \kappa _1+v\right)
   \\
   -\frac{\e^{-\left(\kappa _1+v\right)^2} \left(4 \sqrt{\pi } \e^{2\kappa _1^2} \kappa _1^3 \left(\left(\rho _- -\rho _+\right)^3
   \erfc\left(\sqrt{2} \kappa _1\right)+2 \rho _+ \left(\rho
   _+^2+\rho _- \left(2-3 \rho _+\right)\right)\right)-2 \sqrt{2}
   \kappa _1^2 \left(\rho _--\rho _+\right)^3\right)}{2 \left(\rho _--\rho _+\right)^2}
   \\
   + \frac{1}{2} \sqrt{\pi } \e^{\kappa _1^2} \kappa _1 \left(\rho _+-\rho_-\right) 
   \left(4 \mathrm{T} \left(\sqrt{2} \kappa _1,\frac{\kappa_1+v}{\kappa _1}\right)
   -4 \mathrm{T}\left(2 \kappa _1+v,\frac{v}{2 \kappa_1+v}\right)
   + \erfc\left(\frac{2 \kappa_1+v}{\sqrt{2}}\right) - \erfc\left(\kappa_1\right)\right)
   \:.
\end{multline}
for $v>0$ and $\kappa_1>0$.

\subsubsection{Another observable: the current through the origin}
\label{sec:Current}

We now consider another observable, which is the flux of particles through the origin during a time $t$~\cite{Derrida:2009SI,Derrida:2009aSI}, which we can write as
\begin{equation}
  Q_t = \sum_{r \geq 1}
  \left(
    \eta_r(t) - \eta_r(0)
  \right)
  \:.
\end{equation}
We consider the cumulant generating function
\begin{equation}
  \psi_Q(\lambda) = \sqrt{2t} \: \hat{\psi}_Q = \ln \moy{\e^{\lambda Q_t}}
  \:,
\end{equation}
and the generalised profiles for the current
\begin{equation}
  w_{Q;r}(t) = \frac{\moy{\eta_r \: \e^{\lambda Q_t}}}{\moy{\e^{\lambda Q_t}}}
  \:.
\end{equation}
And for large $t$, we have
\begin{equation}
  w_{Q;r}(t) \underset{t \to \infty}{\simeq} \Phi_Q(v)
    = \sum_{n\geq 0} \Phi_{Q;n}(v) \frac{\lambda^n}{n!}
  \:,
  \quad
  v = \frac{r-\frac{1}{2}}{\sqrt{2t}}
  \:.
\end{equation}
We still define the functions $\Omega_\pm(v)$ as
\begin{equation}
\label{eq:defOmegaQ}
  \Omega_\pm(v) = \mp 2 \hat{\psi}_Q \frac{\Phi_Q'(v)}{\Phi_Q'(0^{\pm})}
  \:.
\end{equation}
It can be checked that these functions again verify the integral equations~(\ref{eq:sm_IntegEqOmPOnly},\ref{eq:sm_IntegEqOmMOnly}), but with $\xi = 0$:
\begin{equation}
  \label{eq:IntegEqOmPJOnly}
    \Omega_+(v) = -\omega_Q \: \e^{-v^2}
    - \omega_Q \int_{-\infty}^0 \dd z \: \Omega_+(z) \: \e^{-(v-z)^2}
    \:,
\end{equation}
\begin{equation}
  \label{eq:IntegEqOmMJOnly}
    \Omega_-(v) = \omega_Q \: \e^{-v^2}
    -\omega_Q  \int_0^{\infty} \dd z \: \Omega_-(z) \: \e^{-(v-z)^2}
    \:,
\end{equation}
which still imply that
\begin{equation}
  \label{eq:PsiCurrentPolyLog}
  \hat{\psi}_Q = - \frac{1}{2 \sqrt{\pi}}
  \mathrm{Li}_{\frac{3}{2}}(-\omega_Q \sqrt{\pi})
  \:.
\end{equation}

In order to obtain the boundary conditions satisfied by $\Phi$, we write the time evolution of the CGF from the master equation,
\begin{equation}
  \partial_t \ln \moy{\e^{\lambda Q_t}}
  = \frac{1}{2} \left[
    (\e^\lambda -1 )
    \frac{\moy{\eta_0(1-\eta_1) \e^{\lambda Q_t}}}{\moy{\e^{\lambda Q_t}}}
    +(\e^{-\lambda} -1 )
    \frac{\moy{\eta_1(1-\eta_0) \e^{\lambda Q_t}}}{\moy{\e^{\lambda Q_t}}}
  \right]
  \:.
\end{equation}
We proceed similarly for the generalized profiles
\begin{equation}
  \partial_t w_{Q;0} = (\e^{-\lambda} - (\e^{-\lambda}-1)w_{Q;0})
  \frac{\partial_t \ln \moy{\e^{\lambda Q_t}}}{\e^{-\lambda} -1}
  + \frac{w_{Q;-1} - w_{Q,0}}{2}
  \:,
\end{equation}
\begin{equation}
  \partial_t w_{Q;1} = (\e^{\lambda} - (\e^{\lambda}-1)w_{Q;1})
  \frac{\partial_t \ln \moy{\e^{\lambda Q_t}}}{\e^{\lambda} -1}
  + \frac{w_{Q;2} - w_{Q;1}}{2}
  \:.
\end{equation}
Taking the hydrodynamic limit, we get at leading order
\begin{equation}
    (\e^\lambda -1 )
    \Phi_Q(0^-) (1-\Phi_Q(0^+))
    +(\e^{-\lambda} -1 )
    \Phi_Q(0^+) (1-\Phi_Q(0^-))
    = 0
    \:,
\end{equation}
\begin{equation}
    \Phi_Q'(0^-) = 2 \hat{\psi}_Q \left( \frac{1}{1-\e^{\lambda}} -  \Phi_Q(0^-) \right)
    \:,
    \quad
    \Phi_Q'(0^+) = -2 \hat{\psi}_Q \left( \frac{1}{1-\e^{-\lambda}} - \Phi_Q(0^+) \right)
    \:.
\end{equation}
These boundary conditions, combined with the solution of the equations~(\ref{eq:IntegEqOmPJOnly},\ref{eq:IntegEqOmMJOnly}) yield
\begin{equation}
\label{eq:sm_OmegaQ}
  \omega_Q \sqrt{\pi} = \rho_- (1-\rho_+) (\e^{\lambda} - 1) + \rho_+ (1-\rho_-) (\e^{-\lambda} - 1)
  \:.
\end{equation}
With~(\ref{eq:PsiCurrentPolyLog}), this allows us to recover the result of Derrida and Gerschenfeld~\cite{Derrida:2009SI,Derrida:2009aSI} on the cumulant generating function $\hat{\psi}$. Additionally, we obtain the profiles $\Phi_{Q;n}(v)$. For instance
\begin{subequations}
  \label{eq:PhiFirstOrdersQ}
  \begin{align}
    \label{eq:Phi0Q}
  \Phi_{Q;1}(v)
  &= \frac{\rho (1-\rho)}{2} \erfc(v) \:,
  \\
    \label{eq:Phi1Q}
  \Phi_{Q;2}(v)
  &= \frac{\rho(1-\rho)(1-2\rho)}{2} \erfc(v) \:,
  \\
    \label{eq:Phi2Q}
  \Phi_{Q;3}(v)
  &= \frac{\rho(1-\rho)(1-3\rho + 3 \rho^2)}{2} \erfc(v)
    - 3\frac{\rho^2(1-\rho)^2}{4} \erfc \left( \frac{v}{\sqrt{2}} \right)^2
    \:.
\end{align}
\end{subequations}

\subsubsection{Another observable: a generalized current}
\label{sec:GenCurr}

For a given position $x$, we consider the generalized current
\begin{equation}
\label{eq:sm_defGenFlux}
  J_t(x) = \sum_{r \geq 1}
  \left(
    \eta_{r+x}(t) - \eta_r(0)
  \right)
\end{equation}
studied in~\cite{Imamura:2017SI,Imamura:2021SI}. It measures the difference between the number of particles on the positive axis at $t=0$ and the number of particles at the right of $x$ at time $t$. In this Section, we show that the cumulant generating function of the generalized flux which has been recently computed in~\cite{Imamura:2017SI,Imamura:2021SI} can be retrieved from our approach. 

One subtlety is that at $t=0$ the observable $J_0(x) \neq 0$, which results in additional difficulties due to the contribution of this (random) initial value. To circumvent this difficulty, we consider the observable $J_t(x_t)$, with $x_t = \lfloor \xi \: \sqrt{2t} \rfloor$ the integer part of $\xi \sqrt{2t}$, which now verifies $J_0(x_0) = 0$. The time evolution of the cumulant generating function of this observable combines two contributions: the jumps of $x_t$ at times $t_n = (n/\xi)^2/2$ and the evolution at $x_t$ fixed between two jumps. Combining these two contributions, it can be shown that, for large $t$,
\begin{equation}
  \label{eq:EvolCGFfluxKirone}
   \partial_t \ln \moy{\e^{\lambda J_t(x_t)}}
  = \frac{1}{2} \left[
    (\e^\lambda -1 )
    \frac{\moy{\eta_x(1-\eta_{x+1}) \e^{\lambda J_t(x_t)}}}{\moy{\e^{\lambda J_t(x_t)}}}
    +(\e^{-\lambda} -1 )
    \frac{\moy{\eta_{x+1}(1-\eta_x) \e^{\lambda J_t(x_t)}}}{\moy{\e^{\lambda J_t(x_t)}}}
  \right]
  + \frac{\xi}{\sqrt{2t}} \ln [1 + (\e^{-\lambda}-1) w_{J;1}(t)]
  \:,
\end{equation}
where we have used that the jumps occur with density $\xi/\sqrt{2t}$, and defined
\begin{equation}
    w_{J;r}(x,t) = \frac{\moy{\eta_{x+r}(t) \e^{\lambda J_t(x_t)}} }{\moy{\e^{\lambda J_t(x_t)} }}
    \:.
\end{equation}
Similarly, we obtain that the profiles satisfy
\begin{multline}
\label{eq:EvoW0GenFlux}
  \partial_t w_{J;0} = (\e^{-\lambda} - (\e^{-\lambda}-1)w_{J;0})
  \frac{\partial_t \ln \moy{\e^{\lambda J_t(x_t)}}
    - \frac{\xi}{\sqrt{2t}} \ln [1 + (\e^{-\lambda}-1) w_{J;1}(t)]}
  {\e^{-\lambda} -1}
  \\
  + \frac{w_{J;-1} - w_{J;0}}{2}
  + \frac{\xi}{\sqrt{2t}} \left(
    \frac{\e^{-\lambda} w_{J;1}}{1+(\e^{-\lambda}-1)w_{J;1}} - w_{J;0}
  \right)
  \:,
\end{multline}
and an analogous expression holds for $w_{J;1}$. For large times, we have the scalings
\begin{equation}
  \ln \moy{\e^{\lambda J_t(x_t)}} \simeq \sqrt{2t} \: \hat{\psi}_J(\lambda,\xi)
  \:,
  \quad
  w_r(t) \underset{t \to \infty}{\simeq}
  \Phi_J \left(v = \frac{r-\frac{1}{2}}{\sqrt{2t}}, \xi, \lambda \right)
  \:,
\end{equation}
which we will denote by $\Phi_J(v)$ for simplicity. Using these scalings, we get from~(\ref{eq:EvolCGFfluxKirone},\ref{eq:EvoW0GenFlux}):
\begin{equation}
  \label{eq:ZeroVelFluxKirone}
  (\e^\lambda -1 )
  \Phi_J(0^-) (1-\Phi_J(0^+))
  +(\e^{-\lambda} -1 )
  \Phi_J(0^+) (1-\Phi_J(0^-))
  = 0
  \:,
\end{equation}
\begin{equation}
  \label{eq:DerZeroMfluxKirone}
  \Phi_J'(0^\pm) = \mp 2 \Psi \left( \frac{1}{1-\e^{\mp \lambda}} - \Phi_J(0^\pm) \right)
  \:,
  \quad
  \Psi = \hat{\psi}_J - \xi \ln[ 1 + (\e^{-\lambda}-1) \Phi_J(0^+)]
  \:.
\end{equation}

If we define
\begin{equation}
\label{eq:DefOmGenFlux}
  \Omega_\pm(v) = \mp 2 \Psi \frac{\Phi_J'(v)}{\Phi_J'(0^\pm)}
  \:,
\end{equation}
the solution of the integral equations~(\ref{eq:sm_IntegEqOmPOnly},\ref{eq:sm_IntegEqOmMOnly}) combined with the boundary conditions gives
\begin{equation}
    \omega_J \sqrt{\pi} \e^{\xi^2} =
    \rho_- (1-\rho_+) (\e^{\lambda} - 1) + \rho_+ (1-\rho_-) (\e^{-\lambda} - 1)
    \:,
\end{equation}
and we obtain for the cumulant generating function of $J_t(x_t)$:
\begin{equation}
  \hat{\psi}_J(\lambda,\xi) = 
  - \frac{1}{2 \sqrt{\pi}} \mathrm{Li}_{\frac{3}{2}}(- \omega_J \sqrt{\pi}) 
  + \xi \ln [1 + (\e^{-\lambda}-1) \Phi_J(0^+)]
  \:,
\end{equation}
which we can combine with the expression of $\Phi_J(0^+)$ obtained by integration of $\Omega_+$ to obtain
\begin{equation}
    \label{eq:sm_CGFgenflux}
    \hat{\psi}_J(\lambda,\xi) = \xi \ln [ 1 + (\e^{-\lambda}-1) \rho_+]
  - \sum_{n \geq 1} \frac{(-\omega_J \sqrt{\pi} \e^{\xi^2})^n}{2n} \left(
    \frac{\e^{-n \xi^2}}{\sqrt{n \pi}} - \xi \erfc(\sqrt{n} \xi)
  \right)
  \:,
\end{equation}
which coincides exactly with the expression obtained in~\cite{Imamura:2017SI,Imamura:2021SI}. This supports the exactness of our main equations~(\ref{eq:sm_IntegEqOmPOnly},\ref{eq:sm_IntegEqOmMOnly}), along with the definition~\eqref{eq:DefOmGenFlux} of $\Omega_\pm$ in this case.

\subsection{Comparison with Imamura et al for the position of the tracer}

 \subsubsection{Obtaining the cumulant generating function of the tracer's position from the generalized current}

We have shown in Section~\ref{sec:GenCurr} that our main equations~(\ref{eq:sm_IntegEqOmPOnly},\ref{eq:sm_IntegEqOmMOnly}) allow to recover the cumulant generating function $\hat{\psi}_J$~\eqref{eq:sm_CGFgenflux} of the generalized current $J_t$~\eqref{eq:sm_defGenFlux} recently obtained by Imamura et al in~\cite{Imamura:2017SI,Imamura:2021SI}. In these papers, the authors deduce from this result the cumulant generating function $\hat{\psi}$ of the position of the tracer as follows.

First, taking a Legendre transform, one gets the distribution of the current $J_t(x)$, which takes the form
\begin{equation}
    \mathbb{P}(J_t(x) = J) 
    \underset{t \to \infty}{\sim}
    \e^{- \sqrt{2t} \: \varphi_J(\xi, j)}
    \:,
    \quad \text{with} \quad
    j = \frac{J}{\sqrt{2t}}
    \:,
    \quad
    \xi = \frac{x}{\sqrt{2t}}
    \:,
\end{equation}
where
\begin{equation}
    \varphi_J(\xi, j) = \chi_*(\xi,j) j - \hat{\psi}_J(\chi_*(\xi,j),\xi)
    \:,
    \quad
    \left. \frac{\partial \hat{\psi}_J}{\partial \chi}  \right|_{\chi=\chi*} = j
    \:,
    \quad
    \hat{\psi}_J(\chi,\xi) = \lim_{t \to \infty} \frac{1}{\sqrt{2t}} \ln \moy{\e^{\chi J_t(x)}}
    \:.
\end{equation}
We have replaced the parameter $\lambda$ in~\eqref{eq:sm_CGFgenflux} by $\chi$ to avoid confusion with the argument $\lambda$ of $\hat{\psi}(\lambda)$. Second, using that the number of particles to the right of the tracer is conserved, the position $X_t$ of the tracer verifies $J_t(X_t) = 0$. The distribution of the position of the tracer can therefore be obtained as
\begin{equation}
    \mathbb{P}(X_t = x) = \mathbb{P}(J_t(x) = 0)
    \underset{t \to \infty}{\sim}
    \e^{- \sqrt{2t} \: \varphi_J(\xi, 0)}
    \:.
\end{equation}
Finally, taking a Legendre transform of $\varphi_J(\xi, 0)$ yields the cumulant generating function $\hat{\psi}$ of $X_t$. Having recovered the expression of $\hat{\psi}_J$~\eqref{eq:sm_CGFgenflux} derived in~\cite{Imamura:2017SI,Imamura:2021SI}, we have therefore the same cumulant generating function $\hat{\psi}$ of $X_t$ as~\cite{Imamura:2017SI,Imamura:2021SI} from the procedure above.

\subsubsection{An alternative parametrization for the cumulant generating function}

Additionally, we have obtained an alternative parametrization of the cumulant generating function of $X_t$, which takes the simple form~\eqref{eq:sm_relPsiomegaPolyLog} in terms of the parameter $\omega$. This parameter is deduced from $\rho$ and $\lambda$ as follows.

Using the relation~\eqref{eq:sm_decoup_2b} between $\Phi'(0^\pm)$ and $\Phi(0^\pm)$, combined with the solutions~(\ref{eq:sm_SolOmPFourier},\ref{eq:sm_SolOmMFourier}) we obtain the relation~\eqref{eq:sm_RelIntOmPhi0} which allows to determine $\Phi(0^+)$, and a similar one for $\Phi(0^-)$. Combining these results with~(\ref{eq:sm_decoup_3b}), we obtain the first equation
\begin{equation}
\label{eq:sm_RelZeroXiOmega}
    2( \cosh \lambda - 1) =
    \frac{\rho (\e^\lambda - 1)^2}{\e^\lambda - \e^{-Z_+(\omega,\xi)}}
    + \frac{\rho (\e^{-\lambda} - 1)^2}{\e^{-\lambda} - \e^{-Z_-(\omega,\xi)}}
    \:,
\end{equation}
which relates $\xi$ and $\omega$ to $\rho$ and $\lambda$.

The second equation needed to fully determine $\xi$ and $\omega$ is the relation $\xi = \frac{\dd \hat{\psi}}{\dd \lambda}$. Using this expression in~\eqref{eq:sm_RelZeroXiOmega}, together with~\eqref{eq:sm_relPsiomegaPolyLog}, we obtain a nonlinear differential equation, whose solution yields $\hat{\psi}(\lambda)$.

A more convenient parametrization can be obtained by using the conservation relation~\eqref{eq:sm_ConsRel} instead of $\xi = \frac{\dd \hat{\psi}}{\dd \lambda}$. Combined with the solutions~(\ref{eq:sm_SolOmPFourier},\ref{eq:sm_SolOmMFourier}), this equation can be rewritten as
\begin{equation}
\label{eq:ConsRelv2}
    \xi = \frac{1}{2} \partial_\xi \ln \abs{
    \left( \e^{\lambda}- \e^{-Z_+(\omega,\xi)} \right)
    \left( \e^{-\lambda}- \e^{-Z_-(\omega,\xi)} \right)
    }
    \:,
\end{equation}
which can be written as an algebraic equation using the expression of $Z_\pm$~\eqref{eq:sm_defZpm}.

To summarize, given $\lambda$ and $\rho$, the parameters $\xi$ and $\omega$ are obtained by solving~\eqref{eq:sm_RelZeroXiOmega} and~\eqref{eq:ConsRelv2}. The cumulant generating function $\hat{\psi}(\lambda)$ is then straightforwardly deduced from~\eqref{eq:sm_relPsiomegaPolyLog}. Furthermore, since the large deviations function $\phi(\xi)$ of the tracer's position is deduced from $\hat{\psi}$ by the Legendre transform~\eqref{eq:RelPhiPsi}, we straightforwardly obtain from this parametrization that $\phi(\xi) = \lambda \xi - \hat{\psi}(\lambda)$. This is an alternative route to the one followed in~\cite{Imamura:2017SI,Imamura:2021SI}. We have checked numerically that the two parametrizations give the same result, validating our approach.

\subsection{Comparison with Derrida et al for the current through the origin}

Combining our intermediate results~(\ref{eq:PsiCurrentPolyLog},\ref{eq:sm_OmegaQ}) on the current through the origin $Q_t$, we obtain the cumulant generating function 
\begin{equation}
    \hat{\psi}_Q(\lambda) = \lim_{t\to \infty} \frac{1}{\sqrt{2t}} \ln \moy{ \e^{\lambda Q_t} }
    = - \frac{1}{2 \sqrt{\pi}}
    \mathrm{Li}_{\frac{3}{2}} \left(
        -\rho_- (1-\rho_+) (\e^{\lambda} - 1) - \rho_+ (1-\rho_-) (\e^{-\lambda} - 1)
    \right)
    \:,
\end{equation}
which is the result of~\cite{Derrida:2009SI,Derrida:2009aSI}. This gives another validation of our approach.

\section{Extension to other single-file systems}

\subsection{Description of single-file systems in terms of two {transport coefficients}}
\label{sec:GenSingFile}

In the language of fluctuating hydrodynamics~\cite{Spohn:1983SI}, a single-file system can be described at large distance and large time by a fluctuating density field $\rho(x, t)$ that is shown to obey the following equation,
\begin{equation}
\partial_t \rho(x, t) = \partial_x\left[D(\rho(x, t)) \partial_x \rho(x, t) + \sqrt{\sigma(\rho(x, t))}\eta(x, t)\right],
\end{equation}
where $\eta(x,t)$ is a normalized Gaussian white noise uncorrelated in space and time. The quantities $D(\rho)$ and $\sigma(\rho)$ were first defined from the microscopic details of a lattice gas~\cite{Spohn:1983SI}. It is nevertheless more intuitive to define them for a system of size $L$ between two reservoirs at densities $\rho_a$ and $\rho_b$~\cite{Derrida_2007SI}.
The number of particles transferred from left to right at time $t$ is denoted by $Q_t$ and is shown to satisfy
\begin{align}
    \lim_{t\to\infty} \frac{\left\langle Q_t\right\rangle}{t}
    &= \frac{D(\rho)}{L}(\rho_a - \rho_b) 
    \text{\qquad if } (\rho_a - \rho_b) \text{ is small,} 
    &
    \lim_{t\to\infty} \frac{\left\langle Q_t^2\right\rangle}{t}
    &= \frac{\sigma(\rho)}{L} 
    \text{\qquad if } \rho_a=\rho_b = \rho.
\end{align}
We list below the expressions of $D(\rho)$ and $\sigma(\rho)$ for a few models considered here.
\[
\renewcommand{\arraystretch}{2}
\begin{array}{l*2{|>{\displaystyle}c}}
\text{Model} & D(\rho)  & \sigma(\rho)\\ \hline
\text{Symmetric exclusion process~\cite{Krapivsky:2015SI}} & D_0 & 2 D_0 \rho(1-\rho)  \\
\text{Hard Brownian particles~\cite{Krapivsky:2015SI}} & D_0 & 2D_0 \rho \\
\text{Kipnis-Marchioro-Presutti~\cite{Zarfaty:2016SI}} & D_0 & \sigma_0 \rho^2\\
\text{Random average process~\cite{Krug:2000SI,Kundu:2016SI}} & \frac{\mu_1}{2 \rho^2} & \frac{1}{\rho} \frac{ \mu_1 \mu_2}{\mu_1 - \mu_2}
\end{array}
\]
$D_0$ is the diffusion coefficient of an individual particle, $\sigma_0 = 2a D_0$ with $a$ the lattice constant of the KMP model~\cite{Zarfaty:2016SI}, and $\mu_k$ are the moments of the probability law of the jumps in the RAP~\cite{Kundu:2016SI}.

\bigskip

We are interested in the position $X_t$ of the tracer, which we can define as~\cite{Krapivsky:2015SIa}
\begin{equation}
  \label{eq:FoncXt0}
  \int_0^{X_t} \rho(x,t) \dd x
  = \int_0^{\infty} \left( \rho(x,t) - \rho(x,0) \right) \dd x
  \:,
\end{equation}
by expressing that the number of particles to the right of the tracer is conserved. We define the associated cumulant generating function and the generalized profiles
\begin{equation}
\label{eq:defwrGenSF}
    \psi(\lambda) = \ln \moy{ \e^{\lambda X_t}}
    \:,
    \quad
    w_r(\lambda,t) = 
    \frac{\moy{\rho(X_t + r,t) \e^{\lambda X_t}}}{ \moy{\e^{\lambda X_t}} }
    \:.
\end{equation}

We will also consider the current through the origin, which is expressed as
\begin{equation}
  \label{eq:FoncQ}
  Q_t
  = \int_0^{\infty} \left( \rho(x,t) - \rho(x,0) \right) \dd x
  \:,
\end{equation}
and the associated profiles
\begin{equation}
\label{eq:defwrGenSFQ}
    \psi_Q(\lambda) = \ln \moy{ \e^{\lambda Q_t}}
    \:,
    \quad
    w_{Q;r}(\lambda,t) = 
    \frac{\moy{\rho(r,t) \e^{\lambda Q_t}}}{ \moy{\e^{\lambda Q_t}} }
    \:.
\end{equation}

\subsection{Mapping the RAP to the KMP model}

The random average process~\cite{10.1214/EJP.v3-28SI,Krug:2000SI,Rajesh:2001SI} consists of particles on an infinite line, placed at positions $x_k(t)$ with initial density $\rho$. The particles are allowed to move to a random fraction of the distance to the next one, either to the left or to the right with rate $\frac{1}{2}$. Only the first two moments $\mu_1$ and $\mu_2$ of the distribution of this random fraction are relevant in the hydrodynamic limit, in which the system is described by the coefficients $D(\rho)$ and $\sigma(\rho)$ given in the table above.

This model can be mapped onto the Kipnis Marchioro Presutti model~\cite{Kipnis:1982SI,Hurtado_2009SI}, which describes a one dimensional lattice where each site contains an energy $\varepsilon_k = x_k - x_{k-1}$~\cite{Cividini:2016SI}. At random times, the total energy of two neighbouring sites is randomly redistributed on these sites. In the hydrodynamic limit, we can replace the discrete index $k$ by a continuous variable $z$ and consider the density of the spacings (or energies) $\nu(z,t)$, which averages to $1/\rho$. This system is then described by $D(\nu) = \mu_1/2$ and $\sigma(\nu) = \mu_1 \mu_2 \nu^2/(\mu_1 - \mu_2)$~\cite{Kundu:2016SI}. The displacement of the tracer particle (initially at the origin) is then given by
\begin{equation}
\label{eq:RelDeplRAPfluxKMP}
    x_0(t) = \int_{-\infty}^0 (\nu(z,t) - \nu(z,0)) \dd z = - Q_t[\nu]
    \:,
\end{equation}
where $Q_t[\nu]$ is the current through the origin in the KMP model. The cumulant generating function of the tracer's position in the RAP is thus directly related to the one of the current in the KMP model:
\begin{equation}
\label{eq:cumulRAPfromKMP}
    \psi^{(\mathrm{RAP})}(\lambda,t) = \ln \moy{ \e^{\lambda x_0(t)}}
    = \ln \moy{ \e^{-\lambda Q_t}} = \psi_Q^{(\mathrm{KMP})}(-\lambda,t)
    \:.
\end{equation}
This connection further extends to the generalized density profiles associated to these two observables. In order to show it explicitly, it is more convenient to use the conditional profiles discussed in Section~\ref{sec:CondProf}. We consider for the RAP the mean density conditioned on the tracer's position
\begin{equation}
    \moy{ \rho(y + x_0(t),t) | x_0(t) = x }
    \underset{t \to \infty}{\simeq}
    \tilde{\Phi}^{(\mathrm{RAP})} \left(v = \frac{y}{\sqrt{t}}, \xi = \frac{x}{\sqrt{t}}\right)
    \:.
\end{equation}
Similarly, for the KMP model, we introduce the mean density conditioned on the value of the current
\begin{equation}
    \moy{ \nu(z,t) | Q_t[\nu] = Q }
    \underset{t \to \infty}{\simeq}
    \tilde{\Phi}_Q^{(\mathrm{KMP})} \left(u = \frac{z}{\sqrt{t}}, q = \frac{Q}{\sqrt{t}} \right)
    \:.
\end{equation}
Since $\nu(z,t)$ is the spacing between the particles labelled by $z$ and $\rho(x,t)$ the density of particles at position $x$, the two conditional profiles are related by
\begin{equation}
\label{eq:RelProfKMPtoRAP}
    \tilde{\Phi}^{(\mathrm{RAP})}(v(u),\xi) = \frac{1}{\tilde{\Phi}_Q^{(\mathrm{KMP})}(u,-\xi)}
    \:,
\end{equation}
where $v(u)$ is the position of the particle labelled by $u$, in the reference frame of the tracer particle. It can be obtained by writing that
\begin{equation}
    x_k(t) - x_0(t) = \int_{0}^k \nu(z,t) \dd z
\end{equation}
from the definition of $\nu(z,t)$. Taking the average, conditioned on $x_0(t) = -Q_t[\nu] = \xi \sqrt{t}$, we obtain
\begin{equation}
\label{eq:ParamKMPtoRAP}
    v(u) = \int_0^u \tilde{\Phi}_Q^{(\mathrm{KMP})}(u',-\xi) \dd u'
    \:.
\end{equation}
Equations~(\ref{eq:RelProfKMPtoRAP},\ref{eq:ParamKMPtoRAP}) give a parametric expression for the conditional profiles of the RAP. An analogous parametrization is given in~\cite{Cividini:2016SI} for the average density in the presence of a biased tracer, but without conditioning (in our case, this would correspond to a flat density profile). Similarly to the demonstration of Section~\ref{sec:CondProf}, we can show that these profiles are equivalent to the joint cumulants generating functions~(\ref{eq:defwrGenSF},\ref{eq:defwrGenSFQ}),
\begin{equation}
    w_x(\lambda,t) = \frac{\moy{\rho(x+x_0(t),t)  \e^{\lambda x_0(t)}}}{ \moy{\e^{\lambda x_0(t)}} }
    \underset{t \to \infty}{\simeq}
    \Phi^{(\mathrm{RAP})}(v,\lambda) = \tilde{\Phi}^{(\mathrm{RAP})}(v,\xi_*)
    \:,
    \quad
    \xi_* = \frac{\dd \hat{\psi}^{(\mathrm{RAP})}}{\dd \lambda}
    \:,
\end{equation}
\begin{equation}
     w_{Q;z}(\lambda,t) =\frac{\moy{\nu(z,t) \e^{\lambda Q_t[\nu]}}}{ \moy{\e^{\lambda Q_t[\nu]}} }
    \underset{t \to \infty}{\simeq}
    \Phi_Q^{(\mathrm{KMP})}(v,\lambda) = \tilde{\Phi}_Q^{(\mathrm{KMP})}(v,\xi_*)
    \:,
    \quad
    \xi_* = \frac{\dd \hat{\psi}_Q^{(\mathrm{KMP})}}{\dd \lambda}
    \:.
\end{equation}
Therefore, we finally have the parametrization
\begin{equation}
\label{eq:ParamRAPfromKMP}
    \Phi^{(\mathrm{RAP})}(v(u),\lambda) = \frac{1}{\Phi_Q^{(\mathrm{KMP})}(u,-\lambda)}
    \:,
    \quad
    v(u) = \int_0^u \Phi_Q^{(\mathrm{KMP})}(u',-\lambda) \dd u'
    \:.
\end{equation}
Expanding this expression in powers of $\lambda$, we can obtain the profiles $\Phi_n^{(\mathrm{RAP})}(v)$ of the RAP from those associated with the flux in the KMP model. This is done in Section~\ref{sec:ProfRAP} below.

\subsection{Modified equations for the GDP-generating function}

All the models discussed in this paper (including the RAP via the mapping to the KMP model) are described by the situation $D(\rho) = D_0$ with $\sigma''(\rho)$ constant and $\sigma(0) = 0$, so we restrict ourselves to this case.

\subsubsection{For the position of the tracer}

At large times, the cumulant generating function of the tracer's position scales as
\begin{equation}
    \psi(\lambda, t) \equi{t\to\infty} \hat{\psi}(\lambda)\sqrt{4 D_0 t}
    \:,
\end{equation}
and the GDP generating function~\eqref{eq:defwrGenSF} as
\begin{equation} \label{eq:sm_scale_w_ext}
    w_r(\lambda, t) \equi{t\to\infty}  \Phi\left(v = \frac{r}{\sqrt{4 D_0 t}}, \lambda\right)
    \:.
\end{equation}
We still define the functions $\Omega_\pm(v)$ as
\begin{equation}
    \Omega_\pm(v) = \mp 2 \hat{\psi} \frac{\Phi'(v)}{\Phi'(0^\pm)}
     \:.
\end{equation}
Our main equation becomes
\begin{equation}
\label{eq:IntegEqOmGen}
    \Omega_\pm(v) = \mp \omega \: \e^{-(v+\xi)^2 + \xi^2} +
    \frac{\sigma''(0)}{4D_0} \omega \int_{\mathbb{R}^{\mp}} \Omega_\pm(z) \: \e^{-(v-z+\xi)^2 + \xi^2} \dd z
    \:,
\end{equation}
with $\xi = \frac{\dd \hat{\psi}}{\dd \lambda}$, and the boundary conditions
\begin{equation}
    \Phi'(0^\pm) \mp \hat{\psi}\frac{\sigma''(0)}{2 D_0} 
    \frac{\Phi(0^{\pm})}{\e^{\mp \frac{\sigma''(0) \lambda}{4 D_0}}-1}
    = 0
    \:,
    \quad
    \frac{2 \sigma'(0) + \sigma''(0) \Phi(0^+)}{2 \sigma'(0) + \sigma''(0) \Phi(0^-)} 
    = \e^{\frac{\sigma''(0) \lambda}{4 D_0}}
    \:.
\end{equation}
The solution of the integral equation~\eqref{eq:IntegEqOmGen} can be easily deduced from~(\ref{eq:sm_SolOmPFourier},\ref{eq:sm_SolOmMFourier}). In particular, we get from the expression of $\Omega_+(0)$ that
\begin{equation}
    \hat{\psi} = \frac{2 D_0}{\sigma''(0)\sqrt{\pi}}
    \mathrm{Li}_{\frac{3}{2}} \left( \frac{\sigma''(0) \sqrt{\pi}}{4 D_0} \omega \right)
    \:.
\end{equation}

\subsubsection{For the current through the origin}

Our results on the tracer's position can be extended to the current through the origin~\eqref{eq:FoncQ}, generalizing the discussion of Section~\ref{sec:Current} to more general single file systems. The cumulant generating function scales as
\begin{equation}
    \psi_Q(\lambda, t) 
    = \ln \moy{ \e^{\lambda Q_t}}
    \equi{t\to\infty} \hat{\psi}_Q(\lambda)\sqrt{4 D_0 t}
    \:,
\end{equation}
and the GDP generating function~\eqref{eq:defwrGenSFQ} as
\begin{equation}
    w_{Q;r}(\lambda, t) \equi{t\to\infty}  
    \Phi_Q\left(v = \frac{r}{\sqrt{4 D_0 t}}, \lambda\right)
    \:.
\end{equation}
Defining again $\Omega_\pm$ as in~\eqref{eq:defOmegaQ}, these functions satisfy~\eqref{eq:IntegEqOmGen} with $\xi=0$ and $\omega$ replaced by $\omega_Q$, which is related to $\hat{\psi}_Q$ by
\begin{equation}
    \hat{\psi}_Q = \frac{2 D_0}{\sigma''(0)\sqrt{\pi}}
    \mathrm{Li}_{\frac{3}{2}} \left( \frac{\sigma''(0) \sqrt{\pi}}{4 D_0} \omega_Q \right)
    \:.
\end{equation}
The boundary conditions now become
\begin{equation}
    \Phi_Q'(0^\pm) = \mp 2 \hat{\psi}_Q
    \left(
        \frac{\sigma'(0)}{2 D_0} \frac{1}{1 - \e^{\mp \frac{\sigma'(0)}{2 D_0} \lambda}}
        + \frac{\sigma''(0)}{4 D_0} \Phi_Q(0^\pm)
    \right)
    \:,
    \quad
    \frac{\Phi_Q(0^+)(2 \sigma'(0) + \sigma''(0) \Phi_Q(0^-))}
    {\Phi_Q(0^-)(2 \sigma'(0) + \sigma''(0) \Phi_Q(0^+))} 
    = \e^{\frac{\sigma'(0)}{2D_0} \lambda}
    \:.
\end{equation}

\subsection{Profiles and cumulants for the KMP model}

Applying the procedure described above in Section~\ref{sec:ExpLambdaSEP} to the case of the KMP model, with $D(\rho) = D_0$ and $\sigma(\rho) = \sigma_0 \rho^2$, we can obtain the cumulants and profiles for this model.

\subsubsection{For the position of the tracer}

For the tracer's position, we obtain for instance at first orders
\begin{equation}
    \kappa_2^{(\mathrm{KMP})} = \frac{\sigma_0}{\sqrt{2 \pi D_0}}
    \:,
    \quad
    \kappa_4^{(\mathrm{KMP})} = \frac{\left(12+\left(3 \sqrt{2}-8\right) \pi \right) \sigma_0^3}
    {4 \sqrt{2} \pi ^{3/2} D_0^{5/2}}
    \:,
\end{equation}
and the associated profiles
\begin{subequations}
  \begin{align}
      \Phi_1^{(\mathrm{KMP})}(v)
      &= \frac{\rho \sigma_0}{4 D_0} \erfc(v)
      \:,
      \\
      \Phi_2^{(\mathrm{KMP})}(v)
      &=
      \frac{\rho \sigma_0^2}{4 D_0^2} \left( \erfc(v) - \frac{2}{\pi} \e^{-v^2} \right)
      \:,
      \\
      \Phi_3^{(\mathrm{KMP})}(v)
      &=
      \frac{\rho \sigma_0^3}{32 D_0^3} 
      \left( 
      2 \left( 1 + \frac{6}{\pi} \right)\erfc(v) 
      - 24\frac{\sqrt{\pi}-v}{\pi^{3/2}} \e^{-v^2} 
      + 3 \erfc\left( \frac{v}{\sqrt{2}} \right)^2
      \right)
      \:.
  \end{align}
\end{subequations}

\subsubsection{For the current through the origin}

In the case of the current, we obtain the general expression of the cumulant generating function
\begin{equation}
\label{eq:CGFfluxKMP}
    \hat{\psi}_Q^{(\mathrm{KMP})}(\lambda) = \frac{D_0}{\sqrt{\pi} \sigma_0}
    \mathrm{Li}_{\frac{3}{2}} \left( \left(\frac{\sigma_0 \rho \lambda}{2 D_0}\right)^2 \right)
    \:,
\end{equation}
which coincides with the one given in~\cite{Derrida:2009aSI}, and also the profiles
\begin{subequations}
\label{eq:ProfKMPflux}
  \begin{align}
      \Phi_{Q;1}^{(\mathrm{KMP})}(v)
      &= \frac{\rho^2 \sigma_0}{4 D_0} \erfc(v)
      \:,
      \\
      \Phi_{Q;2}^{(\mathrm{KMP})}(v)
      &=
      \frac{\rho^3 \sigma_0^2}{4 D_0^2} \erfc(v)
      \:,
      \\
      \Phi_{Q;3}^{(\mathrm{KMP})}(v)
      &=
      \frac{3 \rho^4 \sigma_0^4}{32 D_0^3} 
      \left( 
      2 \erfc(v) 
      + \erfc\left( \frac{v}{\sqrt{2}} \right)^2
      \right)
      \:.
  \end{align}
\end{subequations}

\subsection{Profiles and cumulants for the RAP}
\label{sec:ProfRAP}

We straightforwardly obtain the cumulant generating function $\hat{\psi}^{(\mathrm{RAP})}$ of the position of a tracer in the RAP from the one of the current in the KMP model~\eqref{eq:CGFfluxKMP} via the relation~\eqref{eq:cumulRAPfromKMP}, which yields
\begin{equation}
    \hat{\psi}^{(\mathrm{RAP})}(\lambda)
    = \lim_{t \to \infty} \frac{1}{\sqrt{4 D(\rho) t}} \psi^{(\mathrm{RAP})}(\lambda,t)
    = \frac{\rho (\mu_1-\mu_2)}{2 \mu_2\sqrt{\pi} }
    \mathrm{Li}_{\frac{3}{2}} \left( \left(\frac{\mu_2 \lambda}{\rho(\mu_1-\mu_2)}\right)^2 \right)
    \:.
\end{equation}
Similarly, the profiles for the RAP can be deduced from the one associated with the current in the KMP model~\eqref{eq:ProfKMPflux} by setting $D_0 = \mu_1/2$ and $\sigma_0 = \mu_1 \mu_2/(\mu_1-\mu_2)$, and using the parametrization~\eqref{eq:RelProfKMPtoRAP}. This gives
\begin{subequations}
  \begin{align}
      \Phi_1^{(\mathrm{RAP})}(v)
      =& \frac{\mu_2}{2(\mu_1-\mu_2)} \erfc(v)
      \:,
      \\
      \Phi_2^{(\mathrm{RAP})}(v)
      =&
      \frac{\mu_2^2}{2\pi \rho(\mu_1-\mu_2)^2} \left(
       \pi \erfc(v)^2 
       - 2 \pi \left( 1 + v\frac{\e^{-v^2}}{\sqrt{\pi}} \right) \erfc(v)
       - 2 \e^{-v^2} + 2 \e^{-2v^2}
      \right)
      \:,
      \\
      \nonumber
      \Phi_3^{(\mathrm{RAP})}(v)
      =& \frac{3}{4 \pi^2} \frac{\mu_1^3}{\rho^2 (\mu_1-\mu_2)^3}
      \Bigg(
        \pi^2 \erfc(v)^3 
        - \pi^2 \left(4 +  \frac{2v(3-v^2)}{\sqrt{\pi}} \e^{-v^2} \right) \erfc(v)^2
        \\
        \nonumber
       & + (2 \pi^2 + 2 \pi (3-2v^2) \e^{-2v^2} + 2\pi (2 v^2 + 4 \sqrt{\pi} v -3) \e^{-v^2}) \erfc(v)
       \\
       &+ 2 \sqrt{\pi} v \e^{-3v^2} 
       - 4 (2\pi + \sqrt{\pi} v) \e^{-2v^2}
       + 2(4 \pi + \sqrt{\pi} v) \e^{-v^2}
      \Bigg)
      \:.
  \end{align}
\end{subequations}

\section{Comparison with MFT}
\label{sec:MFT}

We now compare our results with those obtained in the formalism of the Macroscopic Fluctuation Theory~\cite{Bertini:2001SI,Bertini:2002SI,Bertini:2005SI,Bertini:2009SI,Bertini:2015SI}. We are interested in the study of the single-file system at a large time $T$. We introduce a new density $\tilde{\rho}$, constructed from $\rho(x,t)$ introduced in Section~\ref{sec:GenSingFile} by rescaling the time and position:
\begin{equation}
    \rho(x,t) = \tilde{\rho} \left(u = \frac{x}{\sqrt{T}},\tau = \frac{t}{T} \right)
    \:.
\end{equation}
The probability to evolve from an density $\tilde{\rho}_0$ at $t=0$ to a density $\tilde{\rho}(u,1)$ at $t=T$ is given by~\cite{Derrida:2009aSI}:
\begin{equation}
  \mathbb{P}(\tilde{\rho}_0(u) \longrightarrow \tilde{\rho}(u,1))
  = \int \Df [\tilde{\rho}(u,\tau)] \Df[H(u,\tau)] \: \e^{-\sqrt{T} \: S[\tilde{\rho},H]}
  \:,
\end{equation}
where the action $S$ reads
\begin{equation}
  \label{eq:ActS}
  S[\tilde{\rho},H] =
  \int \dd u \int_0^1 \dd \tau
  \left(
    H \partial_\tau \tilde{\rho} + D(\tilde{\rho}) \partial_u \tilde{\rho} \partial_u H
    - \frac{\sigma(\tilde{\rho})}{2} (\partial_u H)^2
  \right)
  \:.
\end{equation}
The distribution of the initial condition $\tilde{\rho}_0$ is
\begin{equation}
  \mathbb{P}[\tilde{\rho}_0] \simeq \ed^{- \sqrt{T} \: F[\tilde{\rho}_0]}
  \:,
\end{equation}
with
\begin{equation}
  \label{eq:ActF}
  F[\tilde{\rho}(u,0)] = \int \dd v \int_{\rho}^{\tilde{\rho}(u,0)} \dd z \: 
  \frac{2 D(z)}{\sigma(z)}( \tilde{\rho}(u,0)-z)
  \:.
\end{equation}
In this formalism, the moment generating function of the tracer's position is given by~\cite{Krapivsky:2015SIa}
\begin{equation}
\label{eq:DefCumulGenFctMFT}
  \moy{ \e^{\lambda X_T} } \simeq \int \Df \tilde{\rho}_0
  \int \Df [\tilde{\rho}(u,\tau)] \Df[H(u,\tau)] \: \e^{-\sqrt{T} \: (S[\tilde{\rho},H] + F[\tilde{\rho}_0] - \lambda Y[\tilde{\rho}])}
  \:,
\end{equation}
where $Y[\tilde{\rho}] = X_T/\sqrt{T}$ is the rescaled position of the tracer, which is deduced from~\eqref{eq:FoncXt0}:
\begin{equation}
  \label{eq:FoncXt}
  \int_0^{Y[\tilde{\rho}]} \tilde{\rho}(u,1) \dd u
  = \int_0^{\infty} \left( \tilde{\rho}(u,1) - \tilde{\rho}(u,0) \right) \dd u
  \:.
\end{equation}
For large $T$, the integral in~\eqref{eq:DefCumulGenFctMFT} is dominated by the minimum of $S+F-\lambda Y$, taken as a function of $(\tilde{\rho},H)$. We denote this minimum $(q,p)$. These functions satisfy the evolution equations~\cite{Krapivsky:2015SIa}
\begin{align}
  \label{eq:MFT_q}
  \partial_\tau q &= \partial_u[D(q) \partial_u q] - \partial_u[\sigma(q)\partial_u p]
  \:,
  \\
  \label{eq:MFT_p}
  \partial_\tau p &= - D(q) \partial_u^2 p - \frac{1}{2}  \sigma'(q) (\partial_u p)^2 
  \:,
\end{align}
with the terminal condition for $p$
\begin{equation}
  \label{eq:MFT_limitP}
  p(u,\tau=1) = B \Theta(u-Y)
  \:,
  \quad B = \frac{\lambda}{q(Y,1)}
  \:,
\end{equation}
and the initial condition for $q$, expressed in terms of $p(u,0)$:
\begin{equation}
  \label{eq:MFT_limitQ}
  p(u,0) = B \Theta(u) + \int_{\rho}^{q(u,0)} \dd r \frac{2 D(r)}{\sigma(r)}
  \:.
\end{equation}
As shown in~\cite{poncet2021generalisedSI}, our generalised density profiles~\eqref{eq:defwrGenSF} can be deduced from the MFT solution $q(u,\tau)$, since
\begin{equation}
  \label{eq:FctIntegW}
  w_r(\lambda,T)
  \simeq \frac{
    \displaystyle
    \int \Df \tilde{\rho}_0
    \int \Df [\tilde{\rho}(u,\tau)] \Df[H(u,\tau)] \: \tilde{\rho}(Y[\tilde{\rho}] + r/\sqrt{T},1) \: 
    \e^{-\sqrt{T} \: (S[\tilde{\rho},H] + F[\tilde{\rho}_0] - \lambda Y_T[\tilde{\rho}])}
  }
  {
    \displaystyle
    \int \Df \tilde{\rho}_0
    \int \Df [\tilde{\rho}(u,\tau)] \Df[H(u,\tau)] \: \ed^{-\sqrt{T} \: (S[\tilde{\rho},H] +
      F[\tilde{\rho}_0] - \lambda Y_T[\tilde{\rho}])}
  }
  \:,
\end{equation}
which yields from a saddle point estimate:
\begin{equation}
  \label{eq:EquivMFTprof}
  w_r(\lambda,T)
  \simeq 
  \Phi \left(v = \frac{r}{\sqrt{4 D_0 T}} , \lambda \right)
  =
  q\left(\sqrt{4D_0}(v+\xi), \tau=1 \right)
  \:,
  \quad
  \xi = \frac{Y[q]}{\sqrt{4D_0}}
\end{equation}
where we have used here the scaling with time introduced above in the case $D(\rho) = D_0$ for consistency, but this relation between $\Phi$ and $q$ also holds for a general $D(\rho)$ (without rescaling the positions with $D_0$). Aside from a few specific cases (such as the hard Brownian particles), the MFT equations cannot be solved analytically for arbitrary $\lambda$. We thus rely both on a perturbative solution and a numerical resolution of these equations to compare with our results.

\subsection{Perturbative expansion in \texorpdfstring{$\lambda$}{lambda} for the SEP}

We come back to the case of the SEP, corresponding to $D(\rho) = 1/2$ and $\sigma(\rho) = \rho(1-\rho)$. The MFT equations can be solved perturbatively by expanding them in powers of the parameter $B$, defined in~\eqref{eq:MFT_limitP}, which appears explicitly in the equations, as
\begin{equation}
\label{eq:ExpMFTB}
    p(u,\tau) = \sum_{n \geq 1}B^n p_n(u,\tau)
    \:,
    \quad
    q(u,\tau) = \rho + \sum_{n \geq 1}B^n q_n(u,\tau)
    \:,
    \quad
    Y = \sum_{n \geq 1} B^n Y_n
    \:.
\end{equation}
This procedure was carried out for the first orders in~\cite{Krapivsky:2015SIa}. The difficulty is then to relate $B$ and $\lambda$, because the solution $q(u,\tau=1)$ is discontinuous at $u=Y$, which makes it impossible to use the definition~\eqref{eq:MFT_limitP}. The relation between these parameters can still be found by treating $B$ and $\lambda$ independently, and minimizing the resulting cumulant generating function with respect to $B$. This procedure was used in~\cite{Krapivsky:2015SIa}. Here, we use a shortcut: since $Y = \sqrt{2} \xi$ with $\xi = \frac{\dd \hat{\psi}}{\dd \lambda}$ and $\hat{\psi}$ has been determined in~\cite{Imamura:2017SI}, we can use this relation to obtain $B$ as a function of $\lambda$ after inversion of~\eqref{eq:ExpMFTB}. The main difficulty is now to compute $q_n(u,\tau)$ at a given order $n$.

The solution for the first two orders has been computed in~\cite{Krapivsky:2015SIa}, and reads
\begin{equation}
    p_1(u,\tau) = \frac{1}{2} \erfc \left( \frac{-u}{\sqrt{2(1-\tau)}} \right)
    \:,
    \quad
    q_1(u,\tau) = \frac{\rho(1-\rho)}{2}
  \left[
    \erfc \left( \frac{-u}{\sqrt{2(1-\tau)}} \right)
    - \erfc \left( \frac{-u}{\sqrt{2\tau}} \right)
  \right]
  \:,
  \quad
  Y_1 = \sqrt{\frac{2}{\pi}}(1-\rho)
  \:,
\end{equation}
\begin{equation}
  p_2(u,\tau) = - Y_1 K(u|1-\tau)
  + \frac{1-2\rho}{8} \erfc \left( \frac{u}{\sqrt{2(1-\tau)}} \right)
  \erfc \left( \frac{-u}{\sqrt{2(1-\tau)}} \right)
  \:,
  \quad
  Y_2 = 0
  \:,
\end{equation}
\begin{equation}
    q_2(u,\tau) = \frac{\rho(1-\rho)(1-2\rho)}{4}
  \left[
    \erfc \left( \frac{u}{\sqrt{2\tau}} \right)
    + 
    \erfc \left( \frac{u}{\sqrt{2\tau}} \right)
    \erfc \left( \frac{u}{\sqrt{2(1-\tau)}} \right)
    - \frac{4 Y_1}{1-2\rho} K(u|1-\tau)
  \right]
  \:,
\end{equation}
where
\begin{equation}
    K(u|\tau) =  \frac{\e^{-\frac{u^2}{2t}}}{\sqrt{2\pi t}}
\end{equation}
is the heat kernel. Starting from order $3$, the resolution becomes more complex. In~\cite{Krapivsky:2015SIa}, the solutions at order $3$ were written in the form
\begin{multline}
  p_3(u,t) =
  \frac{(1-2\rho)^2}{24} \left[
    -\erf \left( \frac{u}{\sqrt{2(1-\tau)}} \right)
    \erfc \left( \frac{-u}{\sqrt{2(1-\tau)}} \right)
    -\frac{12 Y_1}{1-2\rho}K(u|1-\tau)
  \right]
  \erfc \left( \frac{u}{\sqrt{2(1-\tau)}} \right)
  \\
  -\frac{1}{2} \left(
    \frac{Y_1^2 u}{1-\tau} + (2\rho - 1) Y_1
  \right)
  K(u|1-\tau)
  + \tilde{p}_3(u,\tau)
  \:,
\end{multline}
\begin{multline}
  q_3(u,t) =
  \frac{\rho(1-\rho)(1-2\rho)^2}{12}
  \left[
    \erfc \left( \frac{u}{\sqrt{2\tau}} \right)
    -\erfc \left( \frac{u}{\sqrt{2(1-\tau)}} \right)
  \right]
  \\
  + \frac{\rho(1-\rho)}{2} Y_1 K(u|1-\tau)
  \left(
    -\frac{u}{1-\tau} Y_1
    + (1-2\rho)
    - (1-2\rho) \erfc \left( \frac{u}{\sqrt{2\tau}} \right)
  \right)
  + \tilde{q}_3(u,\tau)
  \:,
\end{multline}
where the two additional functions $\tilde{p}_3$ and $\tilde{q}_3$ satisfy the following inhomogeneous heat equations
\begin{equation}
  \partial_\tau \tilde{p}_3 = -\frac{1}{2} \partial_u^2 \tilde{p}_3
  + q_1 (\partial_u p_1)^2
  \:,
\end{equation}
\begin{equation}
  \partial_\tau \tilde{q}_3 = \frac{1}{2} \partial_u^2 \tilde{q}_3
  - \rho(1-\rho) \partial_u^2 \tilde{p}_3
  + \partial_u(q_1^2 \partial_u p_1)
  \:,
\end{equation}
with the boundary conditions
\begin{equation}
  \tilde{p}_3(u,1) = 0
  \:,
  \quad
  \tilde{q}_3(u,0) =
  - \frac{1}{3\rho(1-\rho)}q_1(u,0)^3 + \rho(1-\rho) \tilde{p}_3(u,0)
  \:.
\end{equation}
However, these equations cannot be solved analytically, and were studied numerically in~\cite{Krapivsky:2015SIa} in order to compute the fourth cumulant $\kappa_4$. One can indeed obtain exact integral representations of the solution $\tilde{q}_3(u,\tau)$, in terms of space-time convolutions of the r.h.s. with the heat kernel. This allow for precise numerical estimate of this function. Here, we are only interested in its value at $\tau=1$ because of the relation~\eqref{eq:EquivMFTprof}. Furthermore, because of the expected form of the equation~\eqref{eq:sm_expectedEqPhi}, the combination
\begin{equation}
    \partial_u^2 \tilde{q}_3(u,1) + u \partial_u \tilde{q}_3(u,1)
\end{equation}
should take a simpler form (the difference in the factors with~\eqref{eq:sm_expectedEqPhi} comes from the fact that $u = \sqrt{2} v$). We can thus write an integral representation for this expression, instead of $\tilde{q}_3$, which can be computed numerically with an arbitrary precision for a large number of points ($\sim 500$). Fitting these points with the functions we expect from the equation~\eqref{eq:EqPhiOrder6p}, we find that for $u>0$
\begin{equation}
  \partial_u^2 \tilde{q}_3(u,1)
  + u \partial_u \tilde{q}_3(u,1)
  = \rho^2 (1-\rho)^2 \left( \alpha u
  \e^{-\frac{u^2}{4}} \erfc \left( \frac{u}{2} \right)
  + \beta \e^{-\frac{u^2}{2}}
  \right)
  \:,
\end{equation}
with
\begin{equation}
    \alpha = 0.07052369794...
    \quad \text{and} \quad
    \beta = -0.07957747154...
    \:.
\end{equation}
The coefficients obtained from this fit are extremely stable: they do not change by more that $10^{-12}$ when the interval or the number of points are changed, or when adding other functions to fit with (these functions then get very small coefficients $< 10^{-10}$). Furthermore, we find that
\begin{equation}
    \abs{\alpha - \frac{1}{8 \sqrt{\pi}} } < 10^{-11}
    \quad \text{and} \quad
    \abs{\beta + \frac{1}{4\pi} } < 10^{-11}
    \:.
\end{equation}
Therefore,
\begin{equation}
  \partial_u^2 \tilde{q}_3(u,1)
  + u \partial_u \tilde{q}_3(u,1)
  = \rho^2 (1-\rho)^2 \left( \frac{u}{8 \sqrt{\pi}}
  \e^{-\frac{u^2}{4}} \erfc \left( \frac{u}{2} \right)
  - \frac{\e^{-\frac{u^2}{2}}}{4\pi} 
  \right)
  \:.
\end{equation}
Although we first obtained this result numerically as described here, it can actually be proved from the integral representation of the l.h.s. in terms of a space-time convolution: the spatial integral can be computed using~\cite{Owen:1980SI}, and the remaining time integration reduces to the above result after some manipulations. Unfortunately, this procedure can only be carried explicitly at this order, while the numerical evaluation can be performed at higher orders. Indeed, using this procedure, we also obtained:
\begin{multline}
    \partial_u^2 q_4(u,1)
  + u \partial_u q_4(u,1)
  = \frac{\rho^2 (1-\rho^2)(1-2\rho)}{16 \pi^2} 
  \left(u \sqrt{\pi} \e^{-\frac{u^2}{4}} \erfc \left( \frac{u}{2} \right) - 2 \e^{-\frac{u^2}{2}} \right)
  \\
  - \frac{\rho^2 (1-\rho)^3}{8 \sqrt{2} \pi^{3/2}} \left(
    \sqrt{\pi} (2+u^2) \e^{-\frac{u^2}{4}} \erfc \left( \frac{u}{2} \right)
    - 2 u \e^{-\frac{u^2}{2}}
  \right)
  \:,
\end{multline}
\begin{multline}
    \partial_u^2 q_5(u,1)
  + u \partial_u q_5(u,1)
  = -\frac{(1-\rho)^2 \rho ^2}{16 \pi ^2}
  \left(\pi  (1-2 \rho)^2-(1-\rho) u \left(\sqrt{2 \pi } (1-2\rho)- u(1-\rho) \right)\right)
  \e^{-\frac{u^2}{2}}
  \\
  +\frac{(1-\rho)^2 \rho ^2}{32 \pi ^{3/2}}
     \left(
   \left(u^2+2\right) \left(\sqrt{2 \pi } (1-2 \rho )-(1-\rho) u\right)+\pi(1-2 \rho )^2 u -(1-\rho) \right)
   \e^{-\frac{u^2}{4}} \erfc\left(\frac{u}{2}\right)
  \\
    + \rho^3 (1-\rho)^3 \left(
         \frac{4 \sqrt{2}-6}{48 \pi} \e^{-\frac{u^2}{2}}
  + \frac{1}{12 \sqrt{2} \pi} \e^{-\frac{u^2}{4}} \erfc \left( \frac{u}{2} \right)
  + \frac{u}{16 \sqrt{\pi}} \e^{-\frac{u^2}{4}} \erfc \left( \frac{u}{2} \right)
  - \frac{u}{6 \sqrt{6\pi}} \e^{-\frac{u^2}{6}} \erfc \left( \frac{u}{2\sqrt{3}} \right)
  \right.
  \\
  \left.
  - \frac{u}{6 \sqrt{6\pi}} \e^{-\frac{u^2}{6}} \erfc \left( \frac{u}{\sqrt{3}} \right)
  + \frac{2u}{3\sqrt{6\pi}} \e^{-\frac{u^2}{6}} \: \mathrm{T}\left( \frac{u}{\sqrt{6}}, \sqrt{3} \right)
    \right)
  \:.
\end{multline}

Going back to the profiles $\Phi_n(v)$ using the relation~\eqref{eq:EquivMFTprof}, these expressions coincide perfectly with the results obtained from our integral equations~(\ref{eq:sm_IntegEqOmPOnly},\ref{eq:sm_IntegEqOmMOnly}).

\subsection{Numerical solution at arbitrary \texorpdfstring{$\lambda$}{lambda}}

The MFT equations~(\ref{eq:MFT_q},\ref{eq:MFT_p}) have a forward/backward structure due to the terminal condition on $p$~\eqref{eq:MFT_limitP} and initial condition on $q$~\eqref{eq:MFT_limitQ}. One can still obtain a numerical solution using the scheme described in~\cite{Krapivsky:2012SI}, which we briefly summarize here.
\begin{enumerate}
\item First solve the equation for $q$~(\ref{eq:MFT_q}) using the initial guess $p(u,\tau) = B \Theta(u-Y)$, corresponding to the terminal condition~(\ref{eq:MFT_limitP}) extended to all times.
\item Then solve the equation for $p$~(\ref{eq:MFT_p}) using the newly obtained function $q(u,\tau)$.
\item Iterate the process, replacing each time either $p$ or $q$ by the newly obtained function. After a few iterations ($\sim 3$), the stability of the algorithm can be improved by replacing the functions by a linear  combination of the last two, e.g.,
\begin{equation}
    q(u,\tau) \longleftarrow
    \alpha q_{\mathrm{new}}(u,\tau) + (1-\alpha)
    q_{\mathrm{old}}(u,\tau)
    \:.
  \end{equation}
For instance with $\alpha = 0.75$.
\end{enumerate}
The Heaviside $\Theta$ function must be regularized in order to use the standard methods for solving partial differential equations. We used the following approximation
\begin{equation}
\label{eq:RegulThetaFct}
    \Theta(u) = \frac{1 + \tanh(a u)}{2}
    \:,
\end{equation}
with $a \sim 100$. This regularization causes a small discrepancy between the numerical solution and the exact one near the discontinuity of the function $q(u,1)$.

This algorithm uses explicitly as inputs the values of $Y$ (the position of the tracer) and $B$, which are related in this problem by the conservation relation~\eqref{eq:FoncXt}. This relation is not satisfied for arbitrary values of both $Y$ and $B$. Given $Y$ as an input, we find the corresponding value of $B$ by performing a dichotomy, until relation~\eqref{eq:FoncXt} is verified. The value of the parameter $\lambda$ is deduced from the definition of $B$~\eqref{eq:MFT_limitP}. This gives the solution $q(u,\tau)$ of the problem for a given position of the tracer $Y$, and thus the profiles $\Phi(v)$ from~\eqref{eq:EquivMFTprof}. The plots for different models given in the main text are in excellent agreement with the solution of our main equations~(\ref{eq:sm_IntegEqOmPOnly},\ref{eq:sm_IntegEqOmMOnly}).

\section{Numerical simulations}

\subsection{Symmetric exclusion process}

The simulations of the SEP are performed on a periodic ring of size $N$, with $M=\rho N$ particles at average density $\rho$. The particles are initially placed uniformly at random. The jumps of the particles are implemented as follow: one picks a particle uniformly at random, along with one direction (left and right with equal probabilities). If the chosen particle has no neighbor in that direction, the jump is performed, otherwise it is rejected. In both cases, the time of the simulation is incremented by a random number picked from an exponential distribution of rate $N$.

We keep track of one particle (the tracer) and compute the moments of its displacement and the generalized density profiles. The averaging is performed over $10^8$ simulation.

\subsection{Kipnis-Marchioro-Presutti model}

We consider a periodic lattice of $N = 500$ sites, each one carrying a continuous energy variable $\varepsilon_i > 0$. Initially the energy of each site is picked independently from a Boltzmann distribution at inverse temperature $\rho=1$. At a random time picked from an exponential distribution of rate $N$, we pick uniformly a site $n$ of the lattice. The total energy of sites $n$ and $n+1$ is randomly redistributed between these sites with a uniform distribution. This process is repeated until the final time $t$ is reached.

The position of the tracer (initially chosen as $0$) is defined as the boundary which delimits two regions where the energy is conserved (upon subtracting the flux through the periodic boundary). In the continuous limit, this is equivalent to the definition~\eqref{eq:FoncXt0}.

The generalized density profiles are averaged over $1.2 \cdot 10^9$ simulations.

\subsection{Random-average process}

By construction of the RAP, if the density of the particles is denoted by $\rho$ and if $x$ and $t$ are the spatial and temporal coordinates, the observables depend only on the two rescaled coordinates $z = \rho x$ and $\tau = \rho t$. For this reason, we only consider the RAP at density $\rho = 1$.

The simulations are performed on a periodic ring of length $L = 10000$, with $N=10000$ particles at positions $x_i$. We choose a uniform probability law for the jumps of the particles. The steady state of the RAP is non-trivial, and can be written in terms of the gaps $g_i = x_{i+1}- x_i$ between two particles (with $x_{N+1} = x_1$) as~\cite{Cividini:2016aSI}
\begin{equation}
  P_{N,L}(\{ g_n \} ) \propto \prod_{n=1}^N \frac{1}{\sqrt{g_n}}
  \:
  \delta \left(
    \sum_{n=1}^N g_n - L
  \right)
  \:.
\end{equation}
Denoting $G_n = \sqrt{g_n}$, this corresponds to a uniform distribution of the vector $(G_1,\ldots,G_N)$ on the $N$-dimensional sphere of radius $\sqrt{L}$. This initial condition can be easily implemented by generating $N$ i.i.d. Gaussian random variables $X_i$ with zero mean and unit variance, and computing
\begin{equation}
    g_i = G_i^2 = L \frac{X_i^2}{\displaystyle \sum_{n=1}^N X_n^2}
    \:.
\end{equation}
The observables of interest are then averaged over $8 \cdot 10^6$ simulations.


\bibliographystyle{apsrev4-1}
%

\end{document}